\documentclass{aa}

\usepackage{graphicx}
\usepackage{txfonts}
\usepackage{amsmath}    
\usepackage{amssymb}    
\usepackage{xspace}
\usepackage[dvipsnames]{xcolor}
\usepackage{enumitem}
\usepackage{natbib}
\bibpunct{(}{)}{;}{a}{}{,} 

\usepackage[colorlinks=true,citecolor=blue,linkcolor=Magenta, urlcolor=RubineRed]{hyperref}
\usepackage{float}

\newcommand{\package}[1]{\textsl{#1}}

\newcommand{\gaia}{\textsl{Gaia}\xspace}

\newcommand{\kpc}{\ensuremath{\textrm{kpc}}\xspace}
\newcommand{\kms}{\ensuremath{\textrm{km}~\textrm{s}^{-1}}\xspace}

\newcommand{\feh}{\ensuremath{[\textrm{Fe} / \textrm{H}]}\xspace}

\newcommand{\Vtan}{\textrm{V}_{tan}\xspace}

\newcommand{\Lz}{\ensuremath{L_{z}}\xspace}
\newcommand{\Lperp}{\ensuremath{L_{\perp}}\xspace}
\newcommand{\Etot}{\ensuremath{E_\textrm{tot}}\xspace}

\renewcommand{\deg}{\ensuremath{^{\circ}}}

\usepackage[switch]{lineno}

\begin{document}


\title{ED-2: A cold but not so narrow stellar stream crossing
	the solar neighbourhood}
\titlerunning{A cold but not so narrow stellar stream}

\author{E. Balbinot\inst{1,2}, A. Helmi\inst{1}, T. Callingham\inst{1} T. Matsuno\inst{1}, E. Dodd\inst{1}, T. Ruiz-Lara\inst{1,3}}
\authorrunning{Balbinot et al.}

\institute{Kapteyn Astronomical Institute, University of Groningen, Postbus 800,
	NL-9700AV Groningen, The Netherlands
	\and
	Leiden Observatory, Leiden University, P.O. Box 9513, NL-2300 RA Leiden, The Netherlands
	\and
	Universidad de Granada, Departamento de Física Teórica y del Cosmos, Campus Fuente Nueva, Edificio Mecenas, 18071, Granada, Spain
}

\date{Received \today; accepted ...}

\abstract
{ED-2 is a stellar stream identified as a compact group in integrals-of-motion
space in a local sample of halo stars from the third Gaia data release.}
{We investigate its nature and possible association with known halo
substructures.}
{We explored the current properties of ED-2 members in phase-space and
also analysed the expected distribution via orbit integration.
In addition, we studied the metallicity of ED-2 using APOGEE DR17 and LAMOST
DR8 (and re-calibrated DR3).}
{ED-2 forms a compact group in the $x-z$ (or $R-z$) plane, showing a
pancake-like structure as it crosses the solar neighbourhood.
Dynamically, it is most similar to the globular clusters NGC 3201 and NGC 6101 and to
the stellar streams Ylgr and Phlegethon. However, its orbit is sufficiently
different for none of these objects to likely be the ED-2 progenitor. We also
find ED-2 to be quite metal poor: all of its stars have $\feh \leq -2.42$,
with a median $\feh=-2.60^{+0.20}_{-0.21}$. At this low metallicity, it is
unlikely that ED-2 stems from any known globular cluster. Instead, ED-2 seems to
be in a similar category as the recently discovered Phoenix and C-19 stellar
streams. We find that ED-2 members are scattered across the whole sky, which is
due to its current orbital phase. We predict that as this object moves to its
next apocentre, it will acquire an on-sky morphology that is similar to that of cold stellar
streams. Finally, because ED-2 is nearing pericentre, we predict that additional
members found below the plane will have high radial velocities of close to
$\sim 500~\kms$ in the present-day direction of the globular cluster NGC 6101.}
{} 

\maketitle

\section{Introduction}

Stellar streams can be defined as groups of stars that were stripped from a
parent galaxy or stellar system (e.g. a globular cluster) by their host galaxy.
These objects are of interest because they provide insights into the evolution
and dynamics of galaxies such as the Milky Way \citep[e.g.][]{Kupper:2015,
	Koposov:2010}. In the past, cold stellar streams have typically been identified
through their spatial coherence on the sky \citep[e.g.][]{Ibata:1994,
	Grillmair:2006, Balbinot:2016}, where they form thin bands stars due to their nearly
identical orbits.

Until recently, many of the streams were identified in large photometric
surveys, such as the Sloan Digital Sky Survey \citep[SDSS;]{Abbott:2019}. The
largest of these streams is the Sagittarius stream \citep{Ibata:1994}. The search for stellar streams has continued in the last decades, and fainter
\citep[e.g. Palomar 5; ][]{Odenkirchen:2001} and more distant \citep[e.g.
	Orphan; ][]{Belokurov:2007b} streams were found using the increasingly more abundant
photometric survey data.

The advent of the Gaia mission has allowed the discovery of new streams through
the use of kinematic data. For example, the identification of several large
substructures associated with different accretion events was possible through
their distinct energy and angular momentum distributions  \citep[for a recent
	review, see][]{Helmi:2020}. These structures often do not show the same degree of
spatial cohesion as their more distant counterparts. Nonetheless, using the
most recent data release of Gaia \citep[DR3;][]{GaiaCollaboration:2021d},  lower
overdensities have now been identified \citep{Dodd:2022,Tenachi:2022},
presumably stemming from objects comparable to the progenitors of the more
distant spatially coherent streams \citep{Tenachi:2022}. Furthermore,
algorithms such as STREAMFINDER have been successful at identifying nearby
narrower streams, which is a direct consequence of one of the hypotheses of the algorithm
\citep{Malhan:2018, Ibata:2021}.

We focus on a structure called ED-2, which is
quite compact in the integrals-of-motion space (i.e. energy and angular momenta; IoM
hereafter) and was identified by \citet{Dodd:2022} using Gaia DR3 and a
combination of ground-based surveys. ED-2 forms a retrograde group, and it
occupies a region in IoM that has been associated with the Sequoia accretion
event in the literature \citep{Koppelman:2019a, Myeong:2019}. A small
fraction of its members has been identified by \citet{Yuan:2020} as members of
the Sequoia accretion event. However, \citet{Dodd:2022} reported ED-2 as an
independent group that is compact in velocity as well, more specifically, in
$v_\phi$ and $v_{z}$. This is indicative that this structure is not phase mixed
in the solar neighbourhood.

In this paper, we aim to investigate the origins of ED-2 and study its
metallicity distribution, and we search for additional members of this
stream. The paper is organized as follows. In Sec. 2 we present the discovery
data for ED-2 and investigate possible associations with known globular
clusters (GCs) and other stellar streams. In Sec. 3 we present an approach to
selecting ED-2 members via photometric selections in combination with Gaia DR3
astrometry. In Sec. 4 we discuss possible formation scenarios for ED-2 and present our
general results.

\section{ED-2 in 6D}
\label{sec:6d}

ED-2 was identified in \citet{Dodd:2022} as a highly retrograde structure with a
tight distribution in energy and angular momenta. The authors reported 32 members
with radial velocities within their 2.5 kpc sample. They also find that ED-2
forms a compact cluster in all velocity components, suggesting that this
structure is not phase mixed and that its stars stream together through
the solar neighbourhood.

We expanded the initial membership of 32 stars to 48 by examining the dendrogram
of the ED-2 cluster produced by the first step of the single-linkage-based
method used in \citet{Dodd:2022}. By relaxing the selection
threshold on the dendrogram, it is possible include more stars in the cluster. As
ED-2 is also tightly grouped in velocity space, we removed several stars with
incompatible velocities, and any additional members are very likely to be part
of the structure. This produced a list of 48 high-confidence members.

\citet{Yuan:2020} have found a cluster similar to ED-2 in the very metal-poor
sample (VMP) of \citet{Li:2018}, consisting of nine stars, four of which are in common
with the sample from \citet{Dodd:2022}. \citet{Yuan:2020} considered the cluster
(\textbf{named DTG-5 in their work}) to be associated with Sequoia
\citep{Koppelman:2019a, Myeong:2019}. Based on the high-degree of similarity to
ED-2, and because it is now clear that the latter is independent of Sequoia, we refer to both simply as ED-2 throughout.

\begin{figure}[ht!]
	\begin{centering}
		\includegraphics[width=\linewidth]{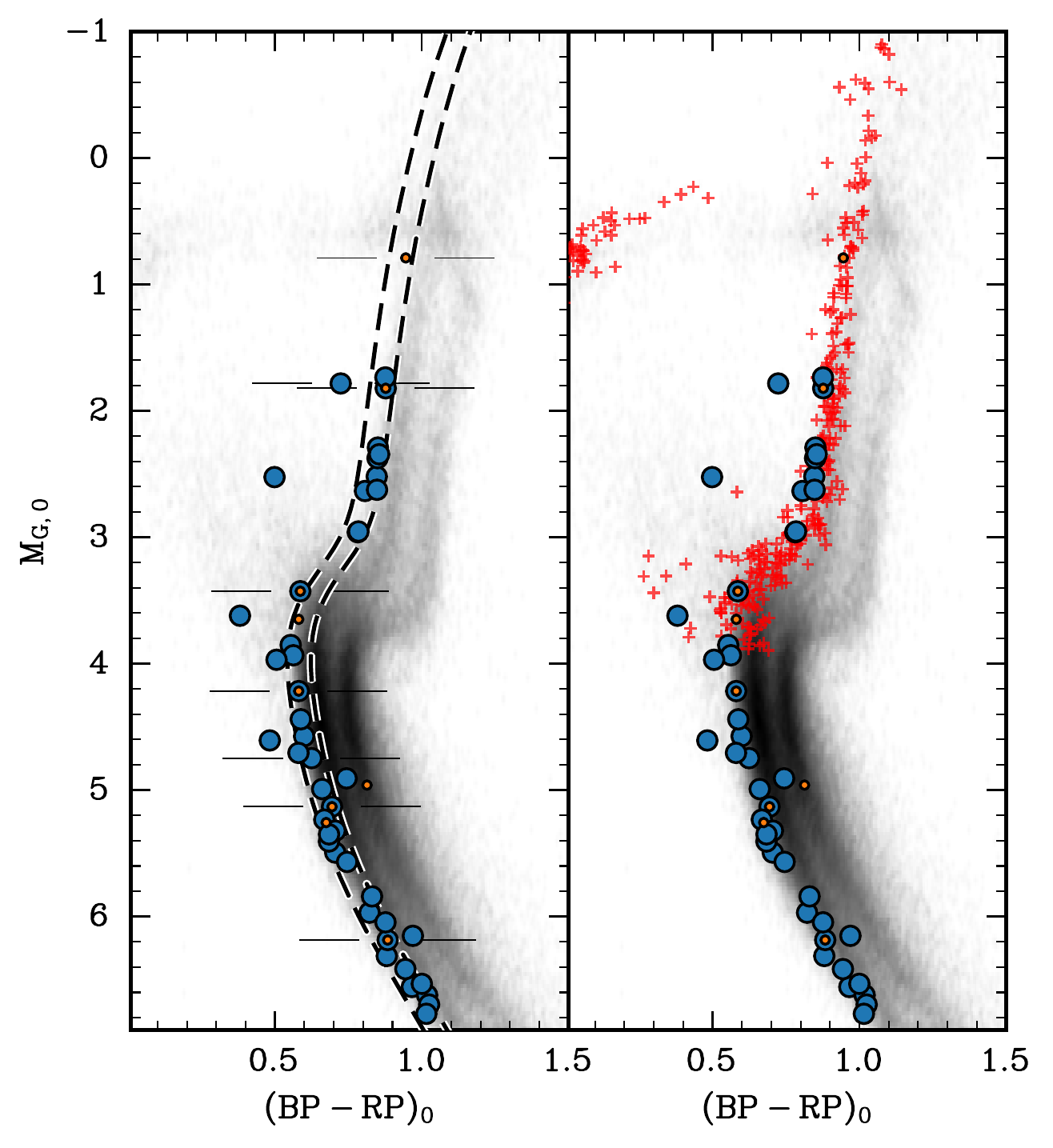}
		\caption{
			Gaia DR3 CMDs showing ED-2 6D members as blue circles.
			In the \textbf{left} panel, the dashed lines were
			constructed using a $\log_{10}{\rm (age/yr)}$ = 10.20
			and \feh=$-2.0$  MIST v1.2 isochrone. In the right panel, we
			show bright, uncrowded, high-probability ($>0.99$)
			members of NGC 6101 \citep{Vasiliev21} with \textbf{red}
			pluses (see Sec. \ref{sec:GC}) In both panels, the
			background shows a linear scale density of halo stars
			selected according to Eq. \ref{eq:halo}. The orange dots
			show the VMP cluster from \citet{Yuan:2020}. In
			the left panel, symbols marked by the horizontal lines
			indicate stars that have spectroscopic metallicities
			(see Fig \ref{fig:spec}).
		}
		\label{fig:CMD}
	\end{centering}
\end{figure}

In Figure~\ref{fig:CMD} we show the members of ED-2 in a colour-magnitude
diagram using Gaia DR3. The distances to ED-2 members were computed by inverting
the parallax after applying the zero-point correction by \citet{Lindegren:2021}.
We note that this distance-estimate method is reliable for distances $\lesssim 5
	\kpc$ and for parallax uncertainties smaller than 20\%, as is the case for ED-2
members. The figure shows that ED-2 members form a very well defined sequence in
the colour-magnitude diagram, and it is consistent with a single stellar
population with a metallicity of $\feh = -2.0$. This estimate is based on the
slope of the red giant branch (RGB) when it is visually compared to a MIST v1.2
isochrone \citep{Choi:2016}. The age of ED-2 is not well constrained because the
main-sequence turn-off (MSTO) is not well populated, but we find that to
simultaneously reproduce the colour of the RGB and main sequence, a
$\log_{10}{\rm (age/yr)} = 10.20$ is necessary. In the same figure, we also show
a MIST isochrone shifted by 0.05 to the blue and 0.03 to the red to define a
photometric selection of additional tentative members (dashed lines).

\footnotetext{We adopted the official Gaia cross match with the survey,
	available at \url{https://gea.esac.esa.int/archive/}.}

In Fig.~\ref{fig:CMD} we also show a sample of stars with
$\varpi/\sigma_{\varpi} > 5$ and $ 1/\varpi < 3$~kpc, with high tangential
velocities, where
\begin{equation}
	\label{eq:halo}
	\Vtan = \frac{k}{\varpi} (\mu_{\alpha^*}^2 + \mu_{\delta}^2)^\frac{1}{2} > 200 \,\kms,
\end{equation}
where $\mu_{\alpha^*, \delta}$ are the proper motion components, $k$ is a
conversion factor, and $\varpi$ is the parallax. This preferentially selects
stars on halo-like orbits \citep[see e.g.][]{GaiaCollaboration:2021c}. We note
that ED-2 members, and as a consequence, the two isochrones we used to identify
additional members, are bluer than the \emph{blue sequence} first reported by \citet[][see also
	\citealt{Koppelman:2019a, GaiaCollaboration:2021c}]{GaiaCollaboration2018b}, and
they are typically associated with the accreted halo component. This is in
favour of ED-2 being more metal poor than the accreted halo. Most ED-2 members
that do not fall within our CMD selection are located in regions of the sky with
E(B-V) $\gtrapprox 0.15$ and/or $|b| < 20\deg$. Hence, these are unlikely to
have reliable colours and magnitudes, regardless of the extinction-correction
prescription adopted. In our case, they were over-corrected for
extinction\footnote{We note that throughout this paper, we corrected all our
	photometric data for extinction using the \citet{Schlegel:1998} dust maps and a
	\citet{Cardelli:1989} extinction curve assuming $R_{V} = 3.1$.}. In the same
figure, we also show (with pluses) a selection of stars from the globular
cluster NGC 6101, which is discussed in more detail in Sec. \ref{sec:GC}.

\begin{figure*}
	\begin{centering}
		\includegraphics[width=\textwidth]{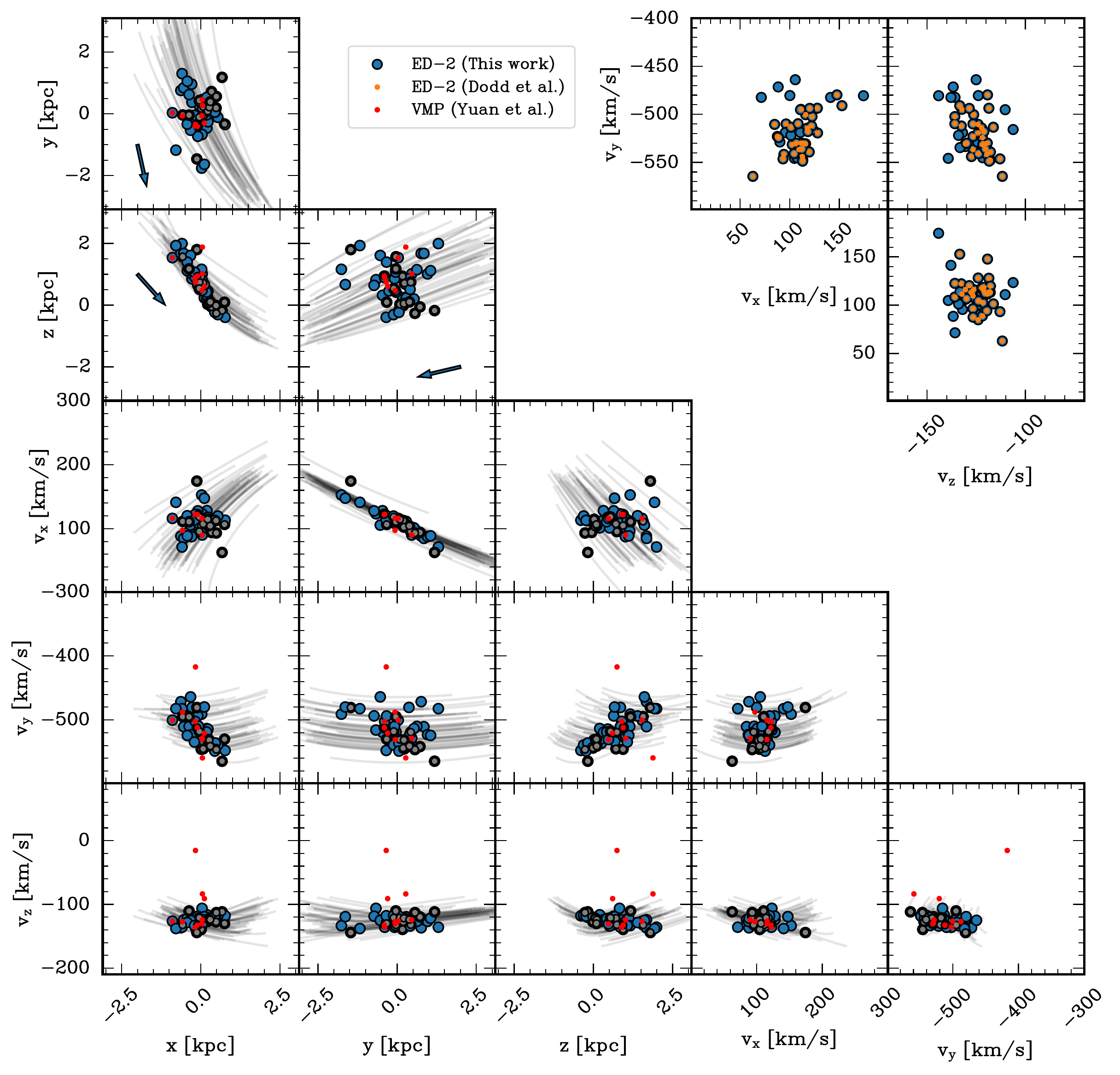}
		\caption{Cartesian heliocentric positions and velocities for ED-2 members.
			The grey symbols mark members that fall outside our narrow photometric selection
			(see Fig. \ref{fig:CMD}). The red dots show the VMP cluster from
			\citet{Yuan:2020}. The lines show a small portion of the orbits of ED-2
			members, integrated in a \textsc{MWPotential} potential for 20 Myr. The
			blue arrow indicates the approximate direction in which the stars move. ED-2 form a tight sequence in the $x-z$ and $v_{x}-y$ panels. We find
			that three objects of \citet{Yuan:2020} are outliers in $v_x$ (outside the plot
			range) and $v_z$ , which are
			unlikely to belong to ED-2. In the three top left panels, the arrows indicate the
			mean projected velocity vector. The three top right panels show
			the original \citet{Dodd:2022} membership (orange) and the expanded membership we
			adopted in this work (blue). We only show these two samples in the
			velocity                                sub-space because this was the sub-space we used to find new
			members of ED-2 (see Sect \ref{sec:6d})
		}
		\label{fig:spatial}
	\end{centering}
\end{figure*}

In Figure \ref{fig:spatial} we show the 3D spatial distribution of ED-2 members
before and after the colour-magnitude selection described above. Additionally,
we highlight the members from the VMP cluster found by \citet{Yuan:2020}. We
note that three of the VMP members disagree with the bulk of ED-2
members and have negative values of $v_x$ and lower values of $v_z$. We mark them as outliers. We find that in the $x-z$ plane, ED-2
forms a tight sequence.
We estimated the stream width by running a principal component analysis
(PCA) decomposition in 3D position space. When computing the standard deviation in each principal axis, we find that ED-2 is distributed in a slightly oval pancake
shape, with dispersions of 750~pc, 650~pc, and 130~pc.
We note that when we propagated the uncertainty in distance, we found that the
typical position uncertainty is smaller than 25 pc. This suggests that ED-2 forms a
pancake-like structure that extends very little spatially  (dispersion of 130~pc), approximately along the
y-axis. We also show in Fig.~\ref{fig:spatial} a small portion of the orbits of ED-2 members, which also reveal preferential alignment in the $x-z$ plane, with a higher density of
overlapping orbits at negative $z$. The orbits were integrated in the
\textsc{MWPotential} potential from \textsc{gala} \citep{gala} for 20 Myr. We
note that the orbit followed by ED-2 members is not exactly aligned with the
stream itself. It resembles an elongated co-moving group of stars more than a
typical stellar stream. This type of morphology-orbit misalignment has been
observed for cold streams that have been perturbed by large satellites
\citep{Shipp:2021}, but a misalignment like this can also arise when a stream approaches a sharp turning point in
its orbit. This seems to be the case for ED-2 because it is close to pericentre (at $\sim$6 kpc) and its velocity vector is aligned with the predicted orbit integrated in the \textsc{MWPotential}.
We find no different orbit or morphology between the CMD selected sample and the
original one from \citet{Dodd:2022}, except that stars near the plane tend to
fall outside our CMD selection. This can be explained by improper extinction
correction, as explained earlier. Finally, we observe a very clear trend in the
$y-v_x$ plane, where ED-2 forms a very tight sequence in velocity. We
performed a PCA analysis in velocity space to determine its dispersion. We find that the lowest velocity dispersion has an amplitude of $\sim$7.4~\kms.
Because the typical velocity uncertainty is $\sim 5 ~\kms$ amongst
ED-2 members and the PCA decomposition does not take any
velocity gradients that might be present into account, the true velocity dispersion of ED-2
is likely lower than our estimate.

\subsection{Metallicity}
\label{sec:met}

\begin{figure}
	\begin{centering}
		\includegraphics[width=\linewidth]{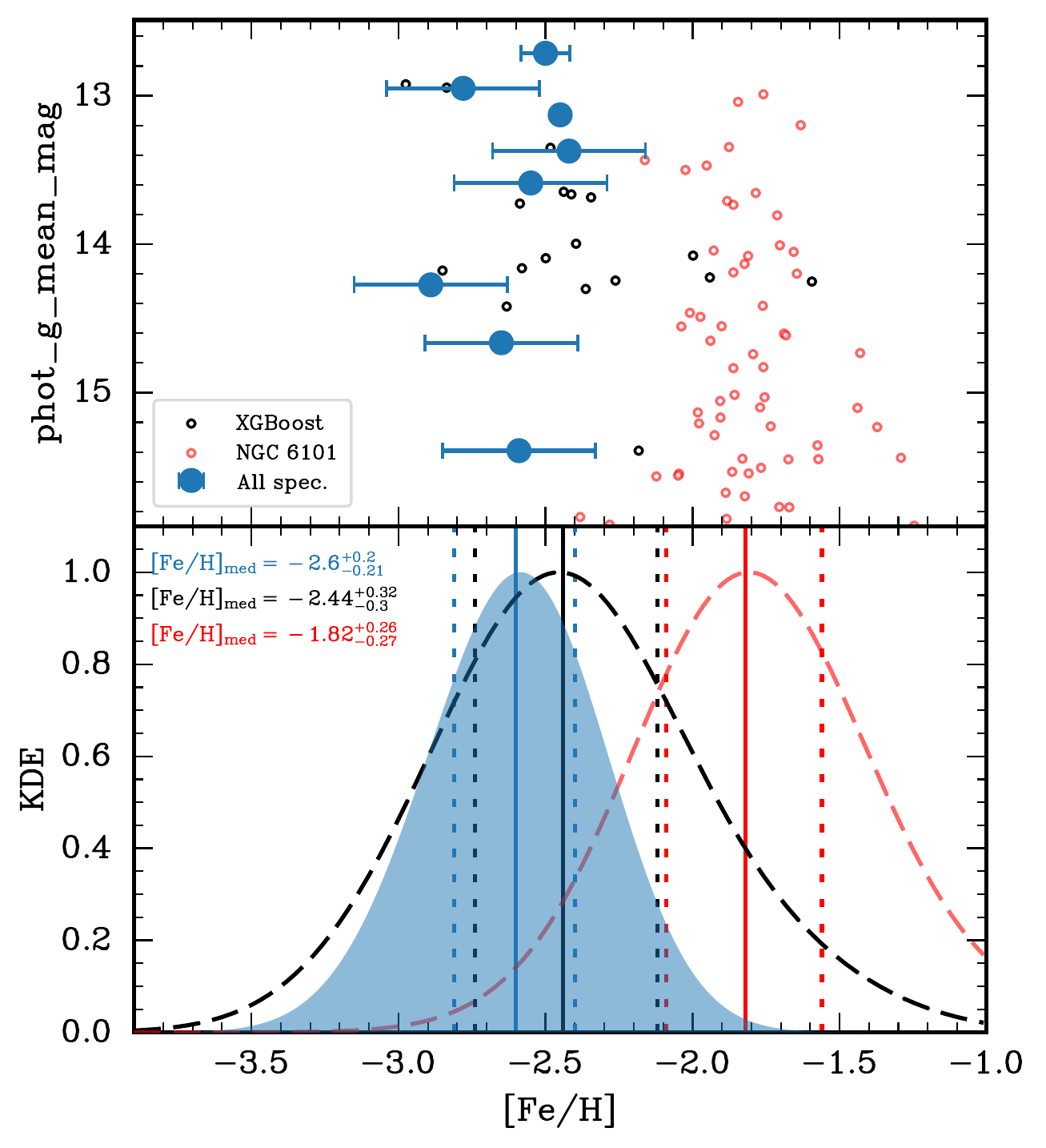}
		\caption{
			Metallicity distribution for ED-2.
			\emph{Top panel:} $\feh$ distribution as a function of observed G-band
			magnitude for the spectroscopic members listed in Table \ref{tab:spec}
			using the \citet{Li:2018VMP} catalogue based on LAMOST and APOGEE data where
			available. We also show the spectro-photometric metallicities from
			\citet{Andrae:2023} after the quality cuts described in Section
			\ref{sec:met}. We show the same for members of NGC 6101. \emph{Bottom
				panel:} KDE of the $\feh$ distribution from spectroscopy and
			\citet{Andrae:2023}. The median (solid) and 25th and 75th percentiles
			(dashed) are shown as vertical lines, matching the colour scheme of the
			two samples in the top panel. We note the good agreement between the
			spectroscopic and spectrophotometric metallicities for ED-2 members.}
		\label{fig:spec}
	\end{centering}
\end{figure}

Some of the ED-2 members are bright enough to have metallicities measured by
ground-based surveys. We find that 22 of these have metallicities, with 21 in
LAMOST DR8 low- and medium-resolution spectra (LRS and MRS) and one in APOGEE DR17. In
Table \ref{tab:spec} we show the summary of the parameters for these targets. We
also list the mean signal-to-noise ratio (S/N) of the spectra. For the case of
LAMOST LRS, the S/N was computed in the $u,g,r,i,z$ bands, and the value in the
table is the highest S/N in any of the bands. The LAMOST DR8
metallicities are shown in parentheses. In the same table, we also
indicate the stars that fall within our CMD selection from Figure \ref{fig:CMD} and
those that are velocity outliers (see Fig. \ref{fig:spatial}).

Additionally, we also took the metallicities from \citet{Li:2018VMP}, who
reanalysed LAMOST DR3 LRS spectra to derive more accurate metallicities for
low-metallicity stars. We find that all but one ED-2 member have metallicities
in this catalogue. We chose to use the \citet{Li:2018VMP} metallicities when
available, as shown in Table \ref{tab:spec}. We note that these rederived
metallicites were shown by \citet{Li:2022} to be reliable when compared to
higher-resolution spectra. However, the authors found that the typical LAMOST
uncertainty were severely underestimated and indicated that an uncertainty of
0.26 dex is more realistic.

In Figure \ref{fig:spec} we show the $\feh$ distribution as a function of
magnitude using the rederived metallicities from \citet{Li:2018VMP}, except for
one star that is not in that catalogue. In the bottom panel, we show a kernel
density estimate (KDE), built by stacking the individual distributions (assumed
to be normally distributed), after removing velocity outliers (see Fig.
\ref{fig:spatial}). The dispersion of each individual measurement was computed
using a fixed value of 0.26 dex (see the discussion above), with the exception of
the single APOGEE stars, for which we used the quoted uncertainty. Finally, we
show the metallicity from Gaia DR3 XP spectra derived via label transfer by
\citet{Andrae:2023} (labelled XGBoost). These metallicities are less
reliable, although for bright (sub-)giants stars, they seem to agree quite well
with the metallicity distributions derived from the spectra. For this latter sample, we
assumed a metallicity uncertainty of 0.35dex. We also adopted a $G < 16$ and $M_{G}
	< 5$ quality cut, which removed low-S/N stars and low-mass main-sequence stars,
which have less reliable spectrophotometric metallicities.

The resulting KDE shows a clear peak, with a median at $\feh =
	-2.60^{+0.20}_{-0.21}$, where the uncertainty is assumed to be the dispersion
around the median computed at the the 25 and 75 percentiles of the cumulative
KDE. We find no difference in the median metallicity when considering only stars
that fall within the CMD selection shown in Figure \ref{fig:CMD}. We also find
that the \citet{Andrae:2023} metallicities agree well with the
spectroscopic ones, with a median $\feh = -2.44^{+0.32}_{-0.30}$.

We note that all the spectroscopic members of ED-2 that have high or medium spectra
and/or have rederived metallicities by \citet{Li:2018VMP}, have $\feh < -2.42$.
This is consistent with our finding based on the CMD distribution, where ED-2
members are found to be bluer than the blue sequence of the halo, which peaks at
a metallicity of $\feh \sim -1.2$ \citep{Conroy:2019}. The median
metallicity of ED-2 is below the metallicity floor of MW GCs \citep{Harris:1996,
	Marin-Franch:2009} and the upper limit of its metallicity only overlaps with a
few GCs, such as Messier 92, NGC 2298, NGC 2419, NGC 4372, NGC 5053, and NGC
6101 \citep[][2010 rev]{Harris:1996}.

\subsection{Association with known structures}
\label{sec:assoc}

\subsubsection*{Globular clusters}
\label{sec:GC}

We investigated possible associations of ED-2 with other known structures in the
Galaxy. First, we sampled 1000 initial conditions for each of the 159 globular
clusters (GCs hereafter) in the \citet[][2022 revision]{Baumgardt18} and
integrated their orbits for 3 Gyr forward and backward in the
\texttt{MWPotential} using \textsc{gala} \citep{gala}.

\begin{figure}
	\begin{centering}
		\includegraphics[width=\linewidth]{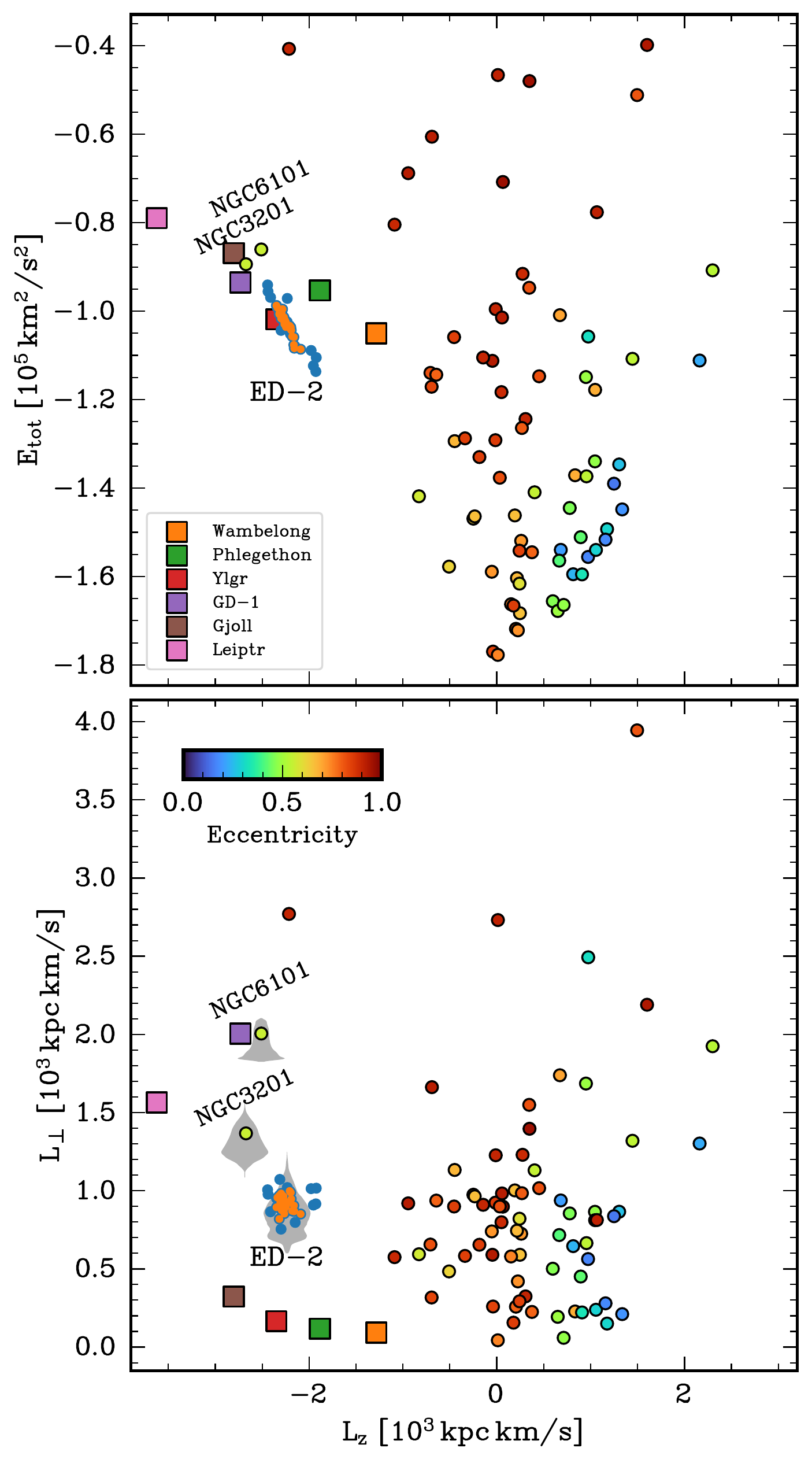}
		\caption{
			ED-2 in IoM space.
			\emph{Top:}
			\Lz vs \Etot showing GCs (circles) coloured by their orbital eccentricity.
			ED-2 is shown as dark blue circles, while the squares show known
			retrograde streams \citep[see the legend;][]{Bonaca:2021}. \emph{Bottom:} Same as
			above, but for \Lz vs \Lperp. Additionally, we show the distribution of
			\Lperp as a violin plot (shaded regions), estimated from the orbits of NGC
			3201, NGC 6101, and ED-2. As a comparison, we also show the
			original ED-2 members identified by \citet{Dodd:2022} (orange circles)
			overlaid on our extended sample (see Sect \ref{sec:6d} and Fig
			\ref{fig:spatial}).
		}
		\label{fig:IoM}
	\end{centering}
\end{figure}

Because of the low degree of phase mixing seen in ED-2, if it is the stellar stream
from a disrupting GC, we expect the progenitor GC to cross the solar
neighbourhood at some point in its orbit. From the simulated GC orbits
described above, we therefore selected only those that approach the Sun closer than 5 \kpc .
We further narrowed down the list of possible progenitors by selecting those that
are nearest to ED-2 in IoM space. The top oanel in Figure \ref{fig:IoM} shows
the $z$-component of the angular momentum ($L_z$) versus total energy ($\Etot$;
computed using the same potential as for the GC orbits) for the GCs in our
sample whose orbits cross the solar neighbourhood. We note that the high-energy
retrograde region of this diagram is typically devoid of GCs, with the exception
of NGC 6101, and NGC 3201. The latter has a somewhat large $\Etot$ uncertainty.
We also show the distribution of $L_{\perp} \equiv (L_x^2 + L_y^2)^{\frac{1}{2}}
$, which is a quantity that is conserved approximately, in the bottom panel of the same
figure. For NGC 3201 and NGC 6101, we show a violin plot in which the width is
proportional to the amount of time for which each cluster orbit has a given value of
$L_{\perp}$. These values were sampled from the orbit simulations described
previously in this section. For ED-2 stars, we show the same, but we did not
sample from the uncertainties of each star, but took the distribution of all ED-2
orbits. From this, we conclude that ED-2 does not overlap with any known GC in
this space.

In addition to the orbital mismatch, NGC 3201 ($\feh = -1.59$) and NGC 6101
($\feh = -1.98$) \citep{Harris:1996} are both more metal rich than ED-2.
Although NGC 3201 shows the highest degree of kinematic association with ED-2,
we conclude from its metallicity that it is unlikely to be the progenitor of
ED-2. This cluster has also been associated with the Gjoll stream
\citep{Hansen:2020}. In the case of NGC 6101, the metallicity is marginally
consistent with ED-2, although it has been poorly studied thus far.
\citet{Dalessandro:2015} reported a  \feh=$-$2.25 based on isochrone fitting,
while spectroscopic estimates indicate a value of $-$1.98 \citep{Zinn:1984,
	Carretta:2009}.  \citet{Geisler:1995} used CaT lines and reported a value of
\feh=$-$1.86\footnote{We note that this metallicity estimate is not on the same
	scale as in \citet{Carretta:2009}}. \citet{Cohen:2011} reported that the RR Lyrae
period distribution and morphology of the horizontal branch (HB) are consistent
with the spectroscopic metallicity estimate.

We further investigated this preliminary assessment by creating a sample of high
membership probability NGC 6101 stars. We adopted the method of
\citet{Vasiliev21}\footnote{https://github.com/GalacticDynamics-Oxford/GaiaTools}
and defined the sample with a membership probability higher than 0.99 and that
are more than 3.6' from the cluster centre (to avoid crowding issues in the Gaia
XP spectra).  In Fig.~\ref{fig:CMD} we show the distribution of these stars in
colour-magnitude diagram with pluses. We adopted the distance modulus and extinction
value from \citet{Dalessandro:2015}. At a first glance, it would seem that ED-2
agrees well with the stellar population of NGC 6101, although the ED2 RGB
stars appear to be slightly bluer. To confirm this, in Fig.~\ref{fig:spec} we
show the \citet{Andrae:2023} metallicity for NGC 6101 members (in red) for
observed G-band magnitudes brighter than 16. All of these stars are RGB stars at
the distance of NGC 6101. We note that the median metallicity of this sample is
$-1.82^{+0.26}_{-0.27}$, and it is significantly offset from both the median
spectroscopic and the metallicity for ED-2 derived in Sec.~\ref{sec:met} using the estimates of
\citet{Andrae:2023}.

\begin{figure}[ht]
	\begin{centering}
		\includegraphics[width=\linewidth]{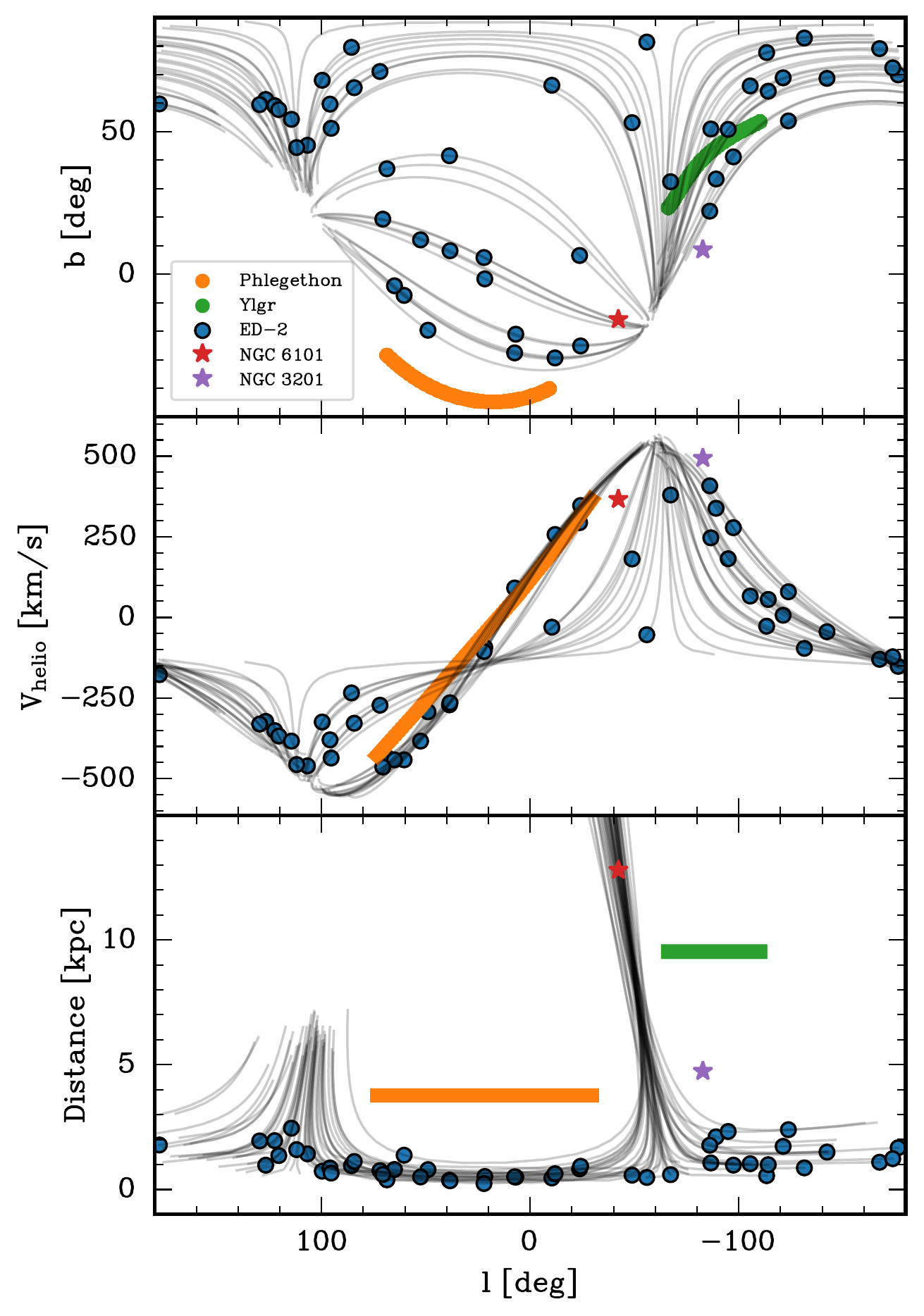}
		\caption{
			Galactic
			$b$, heliocentric velocity ($V_{\rm helio}$), and distance as a function of galactic $l$ from top to bottom. ED-2 members
			are shown in blue, and the grey lines show their predicted orbits. We show
			the Phlegethon and Ylgr streams (see legend) when data are available. The
			red (purple) star marks the location of NGC 6101 (NGC 3201). We note that NGC
			6101 has an excellent match in distance, but only approximately aligns
			with the ED-2 orbit (integrated for 25 Myr and shown in grey) in the other two
			panels.
		}
		\label{fig:stream_association}
	\end{centering}
\end{figure}

Although Fig.~\ref{fig:stream_association} suggests that the expected orbit of
ED-2 crosses the present-day position of NGC 6101 quite well (in distance and on
the sky), its radial velocity is offset by about 100~\kms (\citealt{Harris:1996}
reported 366 \kms for NGC6101), and this mismatch is expected given their
different IoM locations (see Figure \ref{fig:IoM}). Therefore, based on the
dynamical and stellar population differences, we conclude that NGC 6101 is
unlikely to be the progenitor of ED-2.

\subsubsection*{Stellar streams}
\label{sec:streamsorbits}

The census of stellar streams in the MW is rapidly growing
\citep[e.g.][]{Shipp:2018, Ibata:2019}, and currently, more than 100 streams
have been claimed \citep[see][for a recent compilation]{galstreams}. Some of
these streams may cross the solar neighbourhood and may give rise to IoM groups
such as ED-2.

We took a set of retrograde streams from \citet{Bonaca:2021}. We note that not
all these streams have spectroscopic measurements of their radial velocity. Some
are predicted values based on the STREAMFINDER algorithm. Nonetheless, when we
inspect their (sometimes tentative) location of the streams in IoM space, we
find three possible associations with ED-2, namely Phlegethon \citep{Ibata:2018},
Ylgr \citep{Ibata:2019}, and GD-1 \citep{Grillmair:2006}, based on their energy
and $L_z$. We note, however, that they appear to be offset in $L_\perp$ , as shown in Fig.
\ref{fig:IoM}.

GD-1 has been well studied, also spectroscopically, and its orbit is known to
not cross the solar neighbourhood \citep{Koposov:2010}, which is a reflection of
its much larger $L_\perp$ at its energy. Information on Phlegeton is more
limited. \citet{Martin:2022} reported a metallicity of $-1.98\pm0.05$  using
narrow-band photometry in the CaHK band, and \citet{Ibata:2018} predicted the
stream orbit based on two SEGUE DR10 \citep{Yanny:2009} member stars. Fig.~\ref{fig:stream_association} shows that the track of Phlegethon is slightly
offset from that of the ED-2 stars orbits and is located at a larger distance (3.8 kpc
compared to $\sim 1$~kpc at the same sky location). Similarly, although there is
overlap between the track of the Ylgr stream with ED-2 stars, at its sky
position, Ylgr is at a distance of 9.5 kpc \citep{Riley:2020}, whereas ED-2 is
well within 3 kpc. We note that Ylgr has no confirmed spectroscopic members, so
the radial velocity is predicted, and hence, so is it location in IoM space.

Based on the mismatch between distance, velocity, position, and metallicity, we
conclude that neither Phlegethon nor Ylgr (nor GD-1) are likely to be the
progenitor of ED-2. It is nonetheless interesting that these retrograde
structures are all abnormally metal poor. It is tempting to suggest that they
may have been part of a larger group of objects born in a similar environment.

\section{Orbit and predicted velocity}

Unlike the more distant cold streams known thus far in the Galaxy, ED-2 is not a
very narrow stream in comparison. This is partly due to its crossing of the solar
volume.  We can estimate its properties beyond the solar neighbourhood using the
ensemble of orbits of its member stars. Because the ED-2 orbit is nearing
pericentre, we expect that any additional members found at smaller
galactocentric distances will have very high radial velocities. Additionally,
as a stream approaches pericentre, it reaches its maximum density (smallest
physical size), which may make it easier to detect
\citep{Helmi:1999a,Balbinot:2018}.

Because the stream is so close to the Sun, the distribution of the line-of-sight
velocities of its members spans a wide range from $-$400 to 400 \kms.
Fig.~\ref{fig:stream_association} shows that as the stars move away from the
Sun, the predicted line-of-sight velocity reaches values of $\sim 500\,\kms$ for $l
	\sim 100\deg$ and $l \sim -50\deg$ (when the stars lie at distances greater than 5
and 10 kpc from the Sun, respectively).

\begin{figure}[ht!]
	\begin{centering}
		\includegraphics[width=\linewidth]{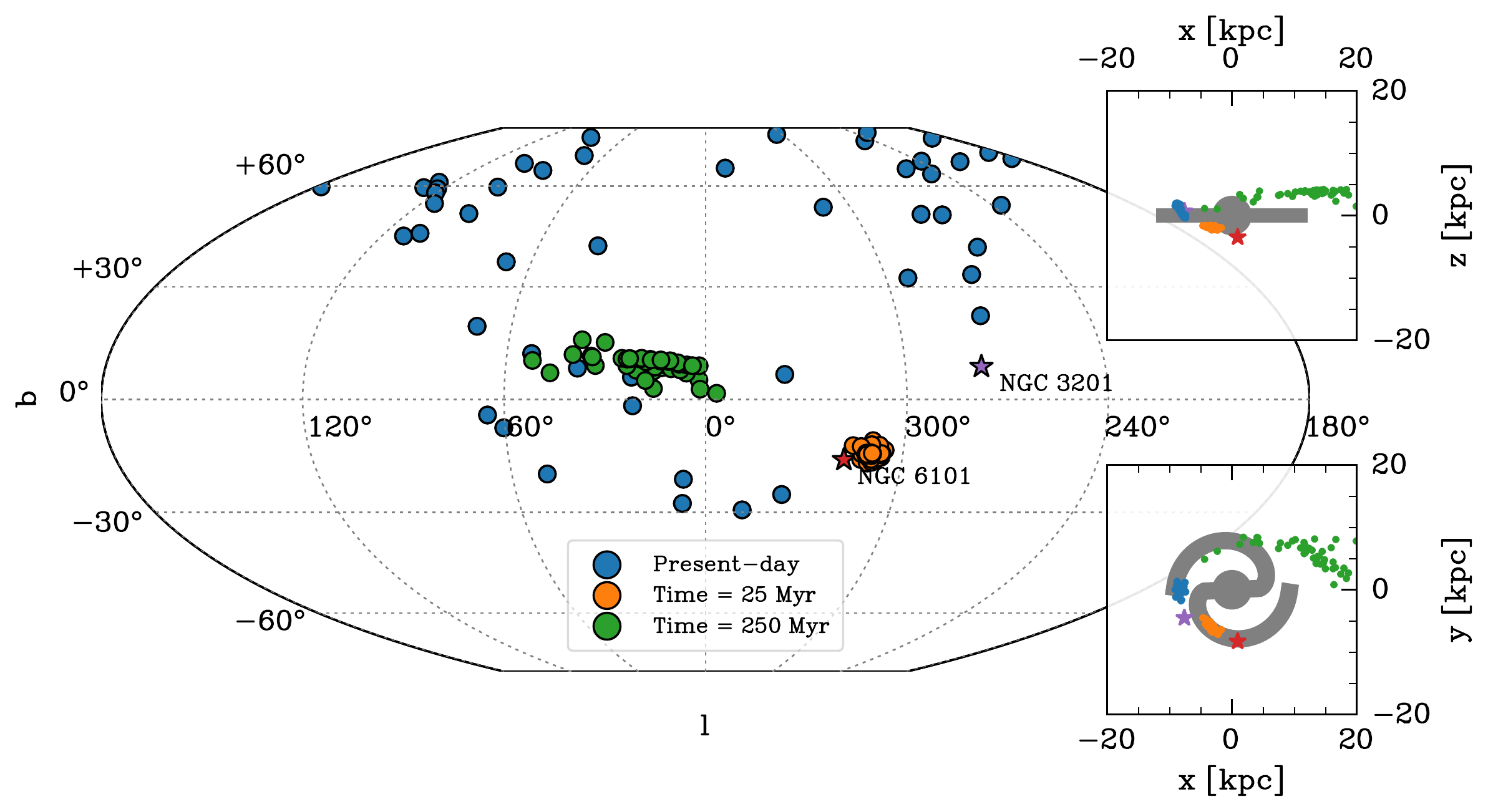}
		\caption{
			All-sky galactic coordinate of the present day (blue), and at 25 (orange), and 250
			(green) Myr in the future position of ED-2 members evolved in a MW-like
			potential. The present-day locations of NGC 6101 and NGC 3201 are also
			shown. In the two insets, we show the $x-y$ and $x-z$ projections. A
			schematic representation of the MW disk and bulge is also given. An
			animated version of this figure is available at
			\url{https://youtu.be/B6UzryhfrHY}
		}
		\label{fig:expected_morph}
	\end{centering}
\end{figure}

To demonstrate how ED-2 morphology would resemble that of a cold stellar stream
when in different orbital phases, we show in Figure~\ref{fig:expected_morph}  the
expected on-sky distribution of its members stars at the present time and at 25 and 250
Myr in the future. We note that initially, ED-2 members converge towards the
present-day position of NGC 6101 (as discussed in Section
\ref{sec:streamsorbits}). However, at its next apocentre, it resembles a thinner
structure in the direction of the MW bulge (but behind it, as shown in the small
insets of Fig.~\ref{fig:expected_morph}).

\begin{table*}
	\footnotesize
	\centering
	\caption{Summary of the ED-2 members with spectroscopic metallicities. The
		dagger                $^{}$ in the designation column indicates whether this target is inside the
		CMD selection from \ref{fig:CMD}, while  $\text{the asterisk}$ shows velocity outliers (see
		Fig. \ref{fig:spatial}). In the last column, we indicate wether this members
		was identified by \citet{Dodd:2022} or \citet{Yuan:2020}.}
	\label{tab:spec}
	\begin{tabular}{crlllrlll}
		\hline
		designation                    & \texttt{source\_id}                              & RA          & Dec        & G      & $\feh$        & $\sigma_{\feh}$ & S/N    & comment   \\
		-                              & -                                                & deg         & deg        & mag    & dex           & dex             & -      & -         \\
		\hline
		                               & \rule{0pt}{2ex} APOGEE DR17 \rule{0pt}{2ex}      &             &            &        &               &                 &        &           \\
		\hline
		2M16242491+2140066             & 1298276602498467072                              & 246.103792  & 21.668526  & 12.714 & -2.50         & 0.083           & 78.82  & Dodd      \\
		\hline
		                               & \rule{0pt}{2ex}LAMOST DR8 MRS   \rule{0pt}{2ex}  &             &            &        &               &                 &        &           \\
		\hline
		$^\dagger$J103946.65+404300.5  & 779616592350301952                               & 159.944397  & 40.7168083 & 13.129 & -2.45         & 0.023           & 34.69  & Dodd/Yuan \\
		\hline
		                               & \rule{0pt}{2ex}LAMOST DR3(8) LRS \rule{0pt}{2ex} &             &            &        &               &                 &        &           \\
		\hline
		$^\dagger$J103946.65+404300.4  & 779616592350301952                               & 159.944389  & 40.716803  & 13.129 & -2.53 (-2.19) & 0.018           & 52.84  & Dodd/Yuan \\
		J112518.93+351846.8            & 759407156314566784                               & 171.328903  & 35.313008  & 12.950 & -2.78 (-2.07) & 0.023           & 57.54  & Dodd      \\
		*$^\dagger$J131903.73+330401.0 & 1472684401070876544                              & 199.765554  & 33.066957  & 16.692 & -2.21 (-1.91) & 0.129           & 23.66  & Yuan      \\
		$^\dagger$J141708.44+444434.6  & 1505084294322771200                              & 214.285167  & 44.7429702 & 15.391 & -2.59 (-2.12) & 0.044           & 57.05  & Yuan      \\
		*J122900.74-000542.7           & 3696558646635083776                              & 187.2530861 & -0.0952172 & 12.884 & -2.15 (-1.82) & 0.078           & 34.00  & Yuan      \\
		J113934.44+101100.5            & 3915925991063287808                              & 174.8935208 & 10.1834806 & 13.588 & -2.55 (-2.14) & 0.027           & 85.61  & Dodd/Yuan \\
		$^\dagger$J125210.47+183309.1  & 3941060277118782208                              & 193.043655  & 18.552555  & 14.664 & -2.65 (-1.88) & 0.065           & 50.53  & Dodd/Yuan \\
		J112352.87+113120.0            & 3963873700285600128                              & 170.970298  & 11.522233  & 14.274 & -2.89 (-2.15) & 0.017           & 143.3  & Dodd/Yuan \\
		*J111719.96+111851.6           & 3963634453427248256                              & 169.333181  & 11.314341  & 14.616 & -2.26 (-1.58) & 0.028           & 108.48 & Yuan      \\
		$^\dagger$J122059.26+245937.2  & 4008274384997266560                              & 185.246941  & 24.993694  & 14.178 & N/A (-1.67)   & 0.097           & 15.61  & Dodd      \\
		$^\dagger$J120727.56+312228.3  & 4014761984637172480                              & 181.864871  & 31.374531  & 13.373 & -2.42 (-2.35) & 0.022           & 137.72 & Yuan      \\


		\hline
	\end{tabular}
\end{table*}


\section{Discussion}

We presented an extensive analysis of the newly discovered ED-2 stream. We find
that the stream forms a tight pancake-like structure, with the least dispersion
in the $x$-$z$ plane. Its stellar population is consistent with an old, single,
metal-poor stellar population with a metallicity of
$\feh=-2.60^{+0.20}_{-0.21}$, based on the reanalysis of ED-2 members in LAMOST
DR3 by \citet{Li:2018}, but also on a high S/N APOGEE member with
$\feh=-2.50\pm0.08$\footnotemark, and a medium-resolution LAMOST DR8 star with
$\feh = -2.45 \pm 0.023$. This estimate is based on a population of six ED-2
members with spectroscopic metallicities that were selected from the original sample of
\citet{Dodd:2022}, with two additional members from \citet{Yuan:2020} that
were selected after velocity outliers were removed. Finally, we find a good
agreement between the spectroscopic metallicity and the Gaia XP derived metallicities
\citep{Andrae:2023}. \footnotetext{We note that the APOGEE DR17 spectral library is
	truncated at $\feh=-2.5$ (see
	\url{https://www.sdss4.org/dr17/irspec/apogee-libraries/}. Thus, the metallicity
	for this star could be an overestimate.}

We closely inspected the possible association with retrograde stellar streams,
namely \textbf{Phlegethon} and Ylgr. We find that despite sharing a similar orbit to
ED-2, the Phlegethon is much more distant and localised on the sky,
additionally its metallicity, if low, is not as low as ED-2. The second
stream, Ylgr, has no radial velocity or metallicity measurements, but we
find that in the regions of the sky in which the expected orbit of ED-2 intersects
the location of Ylgr, there is a large distance mismatch between these two
streams. Other retrograde stellar streams were also considered as progenitors,
but were ruled out because their orbits do not cross the solar
neighbourhood or because previous studies linked them to known progenitors.

The orbit of ED-2 is not consistent with any known GCs, although we find some
similarities with the orbits of NGC 3201 and NGC 6101, clusters that are typically
associated with the Sequoia accretion event \citep{Massari:2019,
	Callingham:2022}. Upon further inspection, we find neither of these GCs to have
fully consistent orbits with the dynamics of ED-2 members. Additionally, we find that
ED-2 is more metal poor than these two GCs. {\it Gaia}-derived metallicities for
high-confidence NGC 6101 members seem to favour a $\feh \sim -1.8$ metallicity
value for this cluster, in agreement with spectroscopic estimates. This
strongly suggests that no known GC is the progenitor of ED-2. Instead, ED-2
seems to be in the same class of progenitor-less stellar streams that have been
recently discovered, such as C-19 \citep{Martin:2022} and the Phoenix stream
\citep{Balbinot:2016, Wan:2020}.

We find that ED-2 resembles a shell or elongated pancake-like structure that is
nearing pericentre. We estimate that ED-2 is a relatively cold stream,
with a thickness of 130 pc and a velocity dispersion of 7.4~\kms. We note
that due to PM and parallax uncertainties, both the size and especially the velocity
dispersion are likely overestimated (given the 5 \kms velocity error typical of its members). The morphology of ED-2 is expected at its
orbital phase, where streams are compressed \citep[and their velocity
	dispersions enhanced; see][]{Helmi:1999a}. From orbital integrations, we find
that ED-2 would resemble a thinner stream if it were observed at apocentre. Based on its
characteristics, a likely scenario is that it had a GC or ultra-faint dwarf
galaxy progenitor.

Recently, \citet{Mikkola:2023} showed that ED-2 is expected to contain roughly
180 members within 3 kpc from the Sun. However, we are lacking radial velocity
measurements to be able to find these additional members. Our orbit
integrations predict a distinct line-of-sight signature beyond the solar
neighbourhood.
In particular, we expect that new members in the direction of NGC 6101 (i.e. closer
to the Galaxy centre and below its plane) have line-of-sight velocities close
to 500 $\kms$.

Future spectroscopic surveys, such as WEAVE and 4MOST, will thus likely allow
the identification of more ED-2 members. Furthermore, higher-resolution
spectroscopic observations of the stream will enable a robust classification of
its progenitor as a GC or ultra-faint dwarf galaxy based on its internal
metallicity spread \citep{Willman:2012} and light-element abundance spreads
\citep{Bragaglia:2017}.

\begin{acknowledgements}
	The authors would like to thank the anonymous reviewer for their comments
	and suggestions that helped improve this manuscript.
	We acknowledge support from a Spinoza prize from the Netherlands
	Organisation for Scientific Research (NWO). The authors would like to thank
	Zhen Yuan for sharing the list of potential ED-2 stars. EB thanks Ivan
	Cabrera-Ziri for the discussion on the chemistry of NGC 6101.
	TRL acknowledges support from Juan de la Cierva fellowship (IJC2020-043742-I),
	financed by MCIN/AEI/10.13039/501100011033. We have made use
	of data from the European Space Agency (ESA) mission \gaia
	(https://www.cosmos.esa.int/gaia), processed by the \gaia Data Processing and
	Analysis Consortium (DPAC,
	https://www.cosmos.esa.int/web/gaia/dpac/consortium).  Funding for the DPAC
	has been provided by national institutions, in particular the institutions
	participating in the \gaia Multilateral Agreement.
	Non-public data underlying this article will be shared on reasonable request to the
	authors. The full list of ED-2 members used in this work is available in the appendix.

	The following software packages where used in this publication:
	\package{Astropy} \citep{astropy, astropy:2018},
	\package{dustmaps} \citep{Green:2018},
	\package{IPython} \citep{ipython},
	\package{matplotlib} \citep{mpl},
	\package{numpy} \citep{numpy},
	\package{scipy} \citep{scipy},
	\package{vaex} \citep{vaex}
\end{acknowledgements}


\bibliographystyle{aa} 
\bibliography{refs.bib} 

\begin{thebibliography}{64}
\expandafter\ifx\csname natexlab\endcsname\relax\def\natexlab#1{#1}\fi

\bibitem[{{Abbott} {et~al.}(2019){Abbott}, {Kron}, {Wang}, {Soergel}, {Nord},
  {Pujol}, {Medford}, { Hollowood}, {March}, {Story}, { Buckley-Geer}, {
  Zenteno}, {Bechtol}, { Zhang}, {Birrer}, {Reil}, {Goldstein}, {Li}, {Davis},
  { D'Andrea}, { Krawiec}, {Sheldon}, {Capozzi}, { Menanteau}, {Lasker},
  {Cawthon}, {Kim}, {Marshall}, {Gangkofner}, {Krause}, { Frieman}, {Ferte },
  {Hamilton}, { Khullar}, {Balbinot}, { Allam}, {Tucker}, {Zhan}, {Vikraman}, {
  Secco}, { Vivas}, {Gerdes}, {Das}, {Kremin }, {Shipp}, {Wiesner}, {Johnson},
  { Bhatawdekar}, {Stebbins}, {Kovacs}, { Bernstein}, {Gelman}, {Sako},
  {Beaufore }, {Wester}, {Elliott}, {Kuropatkin}, { Sheppard}, { Trujillo},
  {Smith}, { Smith}, {Schubnell}, {Clerkin}, { Ogando}, {Annis}, {Valdes}, \&
  {Williams}}]{Abbott:2019}
{Abbott}, T., {Kron}, R., {Wang}, M.~Y., {et~al.} 2019, Minor Planet Electronic
  Circulars, 2019-J52

\bibitem[{{Andrae} {et~al.}(2023){Andrae}, {Rix}, \& {Chandra}}]{Andrae:2023}
{Andrae}, R., {Rix}, H.-W., \& {Chandra}, V. 2023, arXiv e-prints,
  arXiv:2302.02611

\bibitem[{{Astropy Collaboration} {et~al.}(2018){Astropy Collaboration},
  {Price-Whelan}, {Sip{ \H{o}}cz }, {G{\"u}nther}, {Lim}, {Crawford},
  {Conseil}, {Shupe}, {Craig}, { Dencheva}, {Ginsburg}, {VanderPlas}, {Bradley
  }, {P{\'e}rez-Su{ \'a}rez}, {de Val-Borro}, { Aldcroft}, {Cruz},
  {Robitaille}, { Tollerud}, {Ardelean}, {Babej}, {Bach}, {Bachetti},
  {Bakanov}, {Bamford}, { Barentsen}, {Barmby}, {Baumbach}, { Berry},
  {Biscani}, {Boquien}, { Bostroem}, { Bouma}, {Brammer}, {Bray}, {
  Breytenbach}, {Buddelmeijer}, {Burke}, { Calderone}, { Cano Rodr{\' \i}
  guez}, {Cara}, {Cardoso}, { Cheedella}, {Copin}, { Corrales}, { Crichton},
  {D'Avella}, {Deil}, {Depagne }, { Dietrich}, {Donath}, { Droettboom}, {Earl},
  {Erben}, {Fabbro}, { Ferreira}, {Finethy}, {Fox}, {Garrison}, {Gibbons},
  {Goldstein}, {Gommers}, {Greco}, { Greenfield }, {Groener}, {Grollier},
  {Hagen}, {Hirst}, {Homeier}, {Horton}, { Hosseinzadeh}, {Hu}, {Hunkeler}, {
  Ivezi{\'c}}, {Jain}, {Jenness}, { Kanarek}, { Kendrew}, {Kern}, {
  Kerzendorf}, {Khvalko}, {King}, { Kirkby}, {Kulkarni}, {Kumar}, {Lee},
  {Lenz}, {Littlefair}, {Ma}, { Macleod}, { Mastropietro}, {McCully}, {
  Montagnac}, {Morris}, {Mueller}, { Mumford}, {Muna}, { Murphy}, { Nelson},
  {Nguyen }, {Ninan}, {N{ \" o}the}, {Ogaz}, { Oh}, {Parejko}, {Parley},
  {Pascual}, { Patil}, { Patil }, {Plunkett}, {Prochaska}, {Rastogi}, {Reddy
  Janga}, {Sabater}, { Sakurikar}, { Seifert}, {Sherbert}, { Sherwood-Taylor},
  { Shih}, {Sick}, { Silbiger}, {Singanamalla}, {Singer}, {Sladen}, {Sooley},
  {Sornarajah}, {Streicher}, { Teuben}, {Thomas}, {Tremblay}, {Turner},
  {Terr{\'o}n}, {van Kerkwijk}, {de la Vega}, { Watkins}, {Weaver}, {Whitmore},
  {Woillez }, {Zabalza}, \& {Astropy Contributors}}]{astropy:2018}
{Astropy Collaboration}, {Price-Whelan}, A.~M., {Sip{ \H{o}}cz }, B.~M.,
  {et~al.} 2018, \aj, 156, 123

\bibitem[{{Astropy Collaboration} {et~al.}(2013){Astropy Collaboration},
  {Robitaille}, { Tollerud}, {Greenfield}, {Droettboom}, {Bray}, {Aldcroft},
  {Davis}, {Ginsburg}, { Price-Whelan}, {Kerzendorf}, {Conley}, { Crighton},
  {Barbary}, {Muna}, { Ferguson}, {Grollier}, { Parikh}, {Nair}, {Unther},
  {Deil}, {Woillez}, { Conseil}, { Kramer}, {Turner}, {Singer}, {Fox},
  {Weaver}, {Zabalza}, {Edwards}, { Azalee Bostroem}, {Burke}, {Casey},
  {Crawford}, {Dencheva}, {Ely}, {Jenness}, {Labrie}, {Lim}, {Pierfederici},
  {Pontzen}, {Ptak}, {Refsdal}, {Servillat}, \& {Streicher}}]{astropy}
{Astropy Collaboration}, {Robitaille}, T.~P., { Tollerud}, E.~J., {et~al.}
  2013, \aap, 558, A33

\bibitem[{{Balbinot} \& {Gieles}(2018)}]{Balbinot:2018}
{Balbinot}, E. \& {Gieles}, M. 2018, \mnras, 474, 2479

\bibitem[{{Balbinot} {et~al.}(2016){Balbinot}, {Yanny}, {Li}, {Santiago },
  {Marshall}, {Finley}, {Pieres}, { Abbott}, {Abdalla}, {Allam}, {Benoit-L{
  \'e}vy}, {Bernstein}, { Bertin}, {Brooks}, {Burke}, {Carnero Rosell},
  {Carrasco Kind}, {Carretero}, { Cunha}, {da Costa}, { DePoy}, { Desai},
  {Diehl}, {Doel}, { Estrada}, {Flaugher}, {Frieman}, {Gerdes}, {Gruen},
  {Gruendl}, {Honscheid}, { James}, {Kuehn}, {Kuropatkin}, { Lahav}, {March },
  {Martini}, {Miquel}, {Nichol}, { Ogando}, {Romer}, {Sanchez}, { Schubnell},
  {Sevilla-Noarbe}, {Smith}, { Soares-Santos}, {Sobreira}, {Suchyta}, {Tarle},
  {Thomas}, { Tucker}, {Walker}, \& {DES Collaboration}}]{Balbinot:2016}
{Balbinot}, E., {Yanny}, B., {Li}, T.~S., {et~al.} 2016, \apj, 820, 58

\bibitem[{{Baumgardt} \& {Hilker}(2018)}]{Baumgardt18}
{Baumgardt}, H. \& {Hilker}, M. 2018, \mnras, 478, 1520

\bibitem[{{Belokurov} {et~al.}(2007){Belokurov}, {Evans}, {Irwin}, {
  Lynden-Bell}, {Yanny}, {Vidrih}, { Gilmore}, {Seabroke}, {Zucker}, {
  Wilkinson}, { Hewett}, {Bramich}, {Fellhauer}, {Newberg}, {Wyse}, {Beers},
  {Bell}, {Barentine}, {Brinkmann}, {Cole}, {Pan}, \& {York
  }}]{Belokurov:2007b}
{Belokurov}, V., {Evans}, N.~W., {Irwin}, M.~J., {et~al.} 2007, \apj, 658, 337

\bibitem[{{Bonaca} {et~al.}(2021){Bonaca}, {Naidu}, {Conroy}, { Caldwell},
  {Cargile}, {Han}, {Johnson}, {Kruijssen}, { Myeong}, {Speagle}, { Ting}, \& {
  Zaritsky}}]{Bonaca:2021}
{Bonaca}, A., {Naidu}, R.~P., {Conroy}, C., {et~al.} 2021, \apjl, 909, L26

\bibitem[{{Bragaglia} {et~al.}(2017){Bragaglia}, {Carretta}, {D'Orazi}, {
  Sollima}, {Donati}, {Gratton}, \& { Lucatello}}]{Bragaglia:2017}
{Bragaglia}, A., {Carretta}, E., {D'Orazi}, V., {et~al.} 2017, \aap, 607, A44

\bibitem[{{Breddels} \& {Veljanoski}(2018)}]{vaex}
{Breddels}, M.~A. \& {Veljanoski}, J. 2018, \aap, 618, A13

\bibitem[{{Callingham} {et~al.}(2022){Callingham}, {Cautun}, {Deason}, {Frenk},
  {Grand}, \& {Marinacci}}]{Callingham:2022}
{Callingham}, T.~M., {Cautun}, M., {Deason}, A.~J., {et~al.} 2022, \mnras, 513,
  4107

\bibitem[{{Cardelli} {et~al.}(1989){Cardelli}, {Clayton}, \&
  {Mathis}}]{Cardelli:1989}
{Cardelli}, J.~A., {Clayton}, G.~C., \& {Mathis}, J.~S. 1989, \apj, 345, 245

\bibitem[{{Carretta} {et~al.}(2009){Carretta}, {Bragaglia}, {Gratton}, {
  D'Orazi}, \& {Lucatello}}]{Carretta:2009}
{Carretta}, E., {Bragaglia}, A., {Gratton}, R., { D'Orazi}, V., \& {Lucatello},
  S. 2009, \aap, 508, 695

\bibitem[{{Choi} {et~al.}(2016){Choi}, {Dotter}, {Conroy}, {Cantiello},
  {Paxton}, \& {Johnson}}]{Choi:2016}
{Choi}, J., {Dotter}, A., {Conroy}, C., {et~al.} 2016, \apj, 823, 102

\bibitem[{{Cohen} {et~al.}(2011){Cohen}, {Sarajedini}, {Kinemuchi}, \& {
  Leiton}}]{Cohen:2011}
{Cohen}, R.~E., {Sarajedini}, A., {Kinemuchi}, K., \& { Leiton}, R. 2011, \apj,
  727, 9

\bibitem[{{Conroy} {et~al.}(2019){Conroy}, {Naidu}, {Zaritsky}, { Bonaca},
  {Cargile}, {Johnson }, \& { Caldwell}}]{Conroy:2019}
{Conroy}, C., {Naidu}, R.~P., {Zaritsky}, D., {et~al.} 2019, \apj, 887, 237

\bibitem[{{Dalessandro} {et~al.}(2015){Dalessandro}, {Ferraro}, {Massari}, {
  Lanzoni}, {Miocchi}, \& {Beccari}}]{Dalessandro:2015}
{Dalessandro}, E., {Ferraro}, F.~R., {Massari}, D., {et~al.} 2015, \apj, 810,
  40

\bibitem[{{Dodd} {et~al.}(2022){Dodd}, {Callingham}, {Helmi}, { Matsuno},
  {Ruiz-Lara}, { Balbinot}, \& {L{\"o}vdal}}]{Dodd:2022}
{Dodd}, E., {Callingham}, T.~M., {Helmi}, A., {et~al.} 2022, arXiv e-prints,
  arXiv:2206.11248

\bibitem[{{Gaia Collaboration} {et~al.}(2021{\natexlab{a}}){Gaia
  Collaboration}, {Antoja}, {McMillan}, { Kordopatis}, {Ramos}, {Helmi}, {
  Balbinot}, {Cantat-Gaudin}, {Chemin}, { Figueras}, { Jordi}, {Khanna},
  {Romero-G {\'o}mez}, {Seabroke }, {Brown}, {Vallenari}, {Prusti}, {de
  Bruijne}, {Babusiaux}, {Biermann}, {Creevey}, {Evans}, {Eyer}, {Hutton}, {
  Jansen}, {Klioner}, {Lammers}, { Lindegren}, {Luri}, {Mignard}, {Panem}, {
  Pourbaix}, {Randich}, {Sartoretti}, {Soubiran}, {Walton}, {Arenou}, {
  Bailer-Jones}, {Bastian}, {Cropper}, {Drimmel}, {Katz}, {Lattanzi}, { van
  Leeuwen}, {Bakker}, { Casta{\~n}eda}, {De Angeli}, {Ducourant}, { Fabricius},
  {Fouesneau}, {Fr{\'e}mat}, {Guerra}, {Guerrier}, {Guiraud}, {Jean-Antoine
  Piccolo}, {Masana}, {Messineo}, {Mowlavi}, {Nicolas }, {Nienartowicz},
  {Pailler}, { Panuzzo}, {Riclet}, {Roux}, {Sordo}, {Tanga}, {Th{ \'e}venin},
  {Gracia-Abril}, {Portell}, {Teyssier }, {Altmann}, {Andrae},
  {Bellas-Velidis}, {Benson}, {Berthier}, {Blomme}, {Brugaletta}, {Burgess},
  {Busso}, {Carry}, {Cellino }, {Cheek}, {Clementini}, {Damerdji}, { Davidson},
  {Delchambre}, {Dell'Oro}, {Fern{\'a} ndez-Hern{\'a}ndez}, {Galluccio},
  {Garc{\'\i}a-Lario}, {Garcia-Reinaldos}, { Gonz{\'a}lez-N{\'u}{\~n}ez},
  {Gosset}, {Haigron }, {Halbwachs}, { Hambly}, { Harrison}, {Hatzidimitriou},
  { Heiter}, {Hern{\'a}ndez}, {Hestroffer}, {Hodgkin }, {Holl}, {Jan{\ss}en},
  {Jevardat de Fombelle}, {Jordan}, {Krone-Martins}, { Lanzafame },
  {L{\"o}ffler}, {Lorca}, { Manteiga}, {Marchal}, {Marrese}, { Moitinho},
  {Mora}, {Muinonen}, {Osborne}, {Pancino}, {Pauwels}, {Recio-Blanco},
  {Richards}, {Riello}, {Rimoldini}, {Robin}, {Roegiers }, {Rybizki}, { Sarro},
  {Siopis}, { Smith}, { Sozzetti}, {Ulla}, {Utrilla}, { van Leeuwen}, {van
  Reeven}, {Abbas}, {Abreu Aramburu}, {Accart}, {Aerts}, {Aguado}, { Ajaj},
  {Altavilla}, {{\'A}lvarez}, {{ \'A}lvarez Cid-Fuentes}, { Alves}, {Anderson},
  {Varela}, {Audard }, {Baines}, { Baker}, {Balaguer-N{\'u}{\~ n}ez }, {Balog},
  { Barache}, {Barbato}, {Barros }, {Barstow}, {Bartolom{\'e }}, { Bassilana},
  {Bauchet}, { Baudesson-Stella}, {Becciani}, {Bellazzini}, {Bernet},
  {Bertone}, { Bianchi}, { Blanco-Cuaresma}, {Boch}, {Bombrun}, { Bossini}, {
  Bouquillon}, {Bragaglia}, { Bramante}, { Breedt}, {Bressan}, { Brouillet}, {
  Bucciarelli }, {Burlacu}, {Busonero}, { Butkevich}, {Buzzi}, { Caffau}, {
  Cancelliere}, {C{\'a} novas}, {Carballo}, {Carlucci}, { Carnerero},
  {Carrasco}, {Casamiquela}, { Castellani}, { Castro-Ginard}, {Castro Sampol},
  { Chaoul}, {Charlot}, {Chiavassa}, {Cioni}, { Comoretto}, {Cooper}, {
  Cornez}, { Cowell}, {Crifo}, {Crosta}, {Crowley}, { Dafonte}, {Dapergolas},
  {David}, {David}, {de Laverny}, {De Luise}, { De March}, {De Ridder }, {de
  Souza}, {de Teodoro}, {de Torres}, {del Peloso}, {del Pozo}, {Delgado}, {
  Delgado }, { Delisle}, {Di Matteo}, {Diakite}, { Diener}, {Distefano}, {
  Dolding}, { Eappachen }, {Enke}, {Esquej }, {Fabre}, { Fabrizio}, {Faigler},
  {Fedorets}, {Fernique}, { Fienga}, { Fouron}, { Fragkoudi}, {Fraile}, {Franke
  }, {Gai}, {Garabato}, { Garcia-Gutierrez }, {Garc{\' \i}a-Torres},
  {Garofalo}, { Gavras}, { Gerlach}, {Geyer}, {Giacobbe}, {Gilmore}, {Girona},
  {Giuffrida}, { Gomez}, { Gonzalez-Santamaria}, {Gonz{\'a }lez-Vidal},
  {Granvik }, {Guti{\'e}rrez-S{\' a}nchez}, { Guy}, { Hauser}, {Haywood },
  {Hidalgo}, {Hilger}, {H{\l}adczuk }, {Hobbs}, {Holland}, {Huckle},
  {Jasniewicz}, {Jonker}, { Juaristi Campillo }, {Julbe}, {Karbevska}, {
  Kervella}, { Kochoska}, {Kontizas}, { Korn}, { Kostrzewa-Rutkowska},
  {Kruszy{\'n} ska}, { Lambert}, {Lanza}, {Lasne}, {Le Campion}, {Le Fustec},
  {Lebreton }, { Lebzelter}, { Leccia}, {Leclerc}, { Lecoeur-Taibi}, {Liao},
  {Licata}, { Lindstr{\o}m}, {Lister}, { Livanou}, {Lobel}, {Madrero Pardo}, {
  Managau}, { Mann}, {Marchant}, { Marconi}, { Marcos Santos}, {Marinoni},
  {Marocco}, {Marshall}, {Martin Polo}, {Mart{\'\i} n-Fleitas}, { Masip}, {
  Massari}, { Mastrobuono-Battisti}, { Mazeh}, {Messina}, { Michalik},
  {Millar}, {Mints}, { Molina}, {Molinaro}, {Moln{\' a}r}, { Montegriffo}, {
  Mor}, { Morbidelli}, { Morel}, {Morris}, { Mulone}, {Munoz}, { Muraveva}, {
  Murphy}, {Musella}, {Noval}, {Ord{\' e}novic}, {Orr {\`u}}, {Osinde}, {
  Pagani}, {Pagano}, { Palaversa}, { Palicio}, {Panahi}, { Pawlak}, {Pe{\~
  n}alosa Esteller}, {Penttil{\"a} }, { Piersimoni}, {Pineau}, { Plachy},
  {Plum}, {Poggio}, {Poretti}, { Poujoulet }, {Pr{\v {s}}a}, {Pulone},
  {Racero}, {Ragaini}, { Rainer}, { Raiteri}, {Rambaux}, { Ramos-Lerate}, {Re
  Fiorentin}, {Regibo}, {Reyl{ \'e}}, { Ripepi}, { Riva}, {Rixon}, { Robichon},
  {Robin}, {Roelens}, {Rohrbasser}, { Rowell}, {Royer}, {Rybicki}, { Sadowski
  }, { Sagrist{\`a} Sell{\'e}s}, {Sahlmann }, { Salgado}, {Salguero},
  {Samaras}, { Sanchez Gimenez}, {Sanna}, {Santove{\~n}a}, {Sarasso},
  {Schultheis}, {Sciacca}, {Segol}, { Segovia}, {S{\'e}gransan}, {Semeux}, {
  Siddiqui}, {Siebert}, { Siltala}, {Slezak}, {Smart}, {Solano }, {Solitro},
  {Souami}, {Souchay}, {Spagna}, {Spoto}, {Steele}, { Steidelm{\"u}ller},
  {Stephenson}, {S{\"u } veges}, {Szabados}, {Szegedi-Elek}, { Taris},
  {Tauran}, {Taylor}, {Teixeira }, { Thuillot }, {Tonello}, {Torra}, {Torra},
  {Turon}, {Unger}, {Vaillant }, {van Dillen}, {Vanel }, {Vecchiato}, {Viala},
  {Vicente}, { Voutsinas}, {Weiler}, {Wevers}, { Wyrzykowski}, { Yoldas},
  {Yvard}, {Zhao}, {Zorec}, {Zucker}, {Zurbach}, \&
  {Zwitter}}]{GaiaCollaboration:2021c}
{Gaia Collaboration}, {Antoja}, T., {McMillan}, P.~J., {et~al.}
  2021{\natexlab{a}}, \aap, 649, A8

\bibitem[{{Gaia Collaboration} {et~al.}(2018){Gaia Collaboration}, {Babusiaux},
  {van Leeuwen}, { Barstow}, {Jordi}, {Vallenari}, {Bossini}, {Bressan},
  {Cantat-Gaudin}, {van Leeuwen}, { Brown}, {Prusti}, {de Bruijne}, {
  Bailer-Jones}, {Biermann}, {Evans}, { Eyer}, {Jansen }, {Klioner}, {Lammers},
  {Lindegren}, {Luri}, {Mignard}, {Panem}, { Pourbaix}, {Randich},
  {Sartoretti}, { Siddiqui}, {Soubiran}, {Walton}, { Arenou}, { Bastian},
  {Cropper}, {Drimmel }, {Katz}, { Lattanzi}, {Bakker}, {Cacciari}, {Casta{\~n}
  eda}, {Chaoul}, {Cheek}, {De Angeli}, { Fabricius}, {Guerra}, {Holl},
  {Masana}, { Messineo}, {Mowlavi}, {Nienartowicz}, {Panuzzo}, {Portell},
  {Riello}, {Seabroke}, { Tanga}, {Th{\'e}venin}, {Gracia-Abril}, { Comoretto},
  {Garcia-Reinaldos}, {Teyssier }, { Altmann}, {Andrae}, {Audard},
  {Bellas-Velidis}, {Benson}, {Berthier}, {Blomme}, {Burgess}, {Busso}, {
  Carry}, {Cellino}, { Clementini}, { Clotet}, {Creevey}, {Davidson}, {De
  Ridder}, {Delchambre}, {Dell'Oro}, { Ducourant},
  {Fern{\'a}ndez-Hern{\'a}ndez}, { Fouesneau}, {Fr{\'e}mat}, {Galluccio}, {
  Garc{\'\i}a-Torres}, {Gonz{\'a}lez-N{\'u}{\~n}ez}, {Gonz{\'a}lez-Vidal},
  {Gosset}, {Guy}, {Halbwachs}, { Hambly}, { Harrison}, {Hern{\'a}ndez}, {
  Hestroffer}, {Hodgkin}, {Hutton}, {Jasniewicz }, {Jean-Antoine-Piccolo},
  {Jordan}, {Korn}, {Krone-Martins}, {Lanzafame}, { Lebzelter}, {L{\"o}ffler},
  {Manteiga}, { Marrese}, { Mart{\'\i}n-Fleitas}, { Moitinho}, {Mora}, {
  Muinonen}, {Osinde}, {Pancino}, {Pauwels}, {Petit}, {Recio-Blanco},
  {Richards}, {Rimoldini}, {Robin}, {Sarro}, {Siopis}, {Smith}, {Sozzetti},
  {S{\"u}veges}, { Torra}, {van Reeven}, {Abbas}, {Abreu Aramburu}, {Accart},
  {Aerts}, {Altavilla }, {{\'A} lvarez}, {Alvarez}, {Alves}, {Anderson},
  {Andrei}, {Anglada Varela}, {Antiche}, {Antoja}, {Arcay}, {Astraatmadja},
  {Bach}, {Baker}, {Balaguer-N{\'u}{\~n}ez}, {Balm}, { Barache}, {Barata},
  {Barbato}, {Barblan}, {Barklem}, {Barrado}, {Barros}, { Bartholom{\'e}
  Mu{\~n}oz}, {Bassilana}, {Becciani}, {Bellazzini}, {Berihuete}, {Bertone}, {
  Bianchi}, {Bienaym{\'e}}, {Blanco-Cuaresma}, { Boch}, {Boeche}, {Bombrun},
  {Borrachero}, {Bouquillon}, {Bourda}, {Bragaglia}, {Bramante}, { Breddels},
  {Brouillet}, {Br{\"u}semeister}, {Brugaletta}, {Bucciarelli}, {Burlacu}, {
  Busonero}, {Butkevich}, {Buzzi}, {Caffau}, {Cancelliere}, {Cannizzaro},
  {Carballo}, { Carlucci}, {Carrasco}, {Casamiquela}, { Castellani}, {
  Castro-Ginard}, {Charlot}, {Chemin }, { Chiavassa}, {Cocozza}, {Costigan}, {
  Cowell}, {Crifo}, {Crosta}, {Crowley}, { Cuypers}, {Dafonte}, {Damerdji},
  {Dapergolas}, {David}, {David}, {de Laverny}, {De Luise }, {De March}, {de
  Martino}, {de Souza}, {de Torres}, { Debosscher}, {del Pozo}, {Delbo}, {
  Delgado}, {Delgado}, {Diakite}, { Diener}, {Distefano}, {Dolding}, {
  Drazinos}, {Dur{\'a}n}, {Edvardsson}, { Enke}, {Eriksson }, {Esquej}, {Eynard
  Bontemps}, {Fabre}, {Fabrizio}, {Faigler }, {Falc{\~a}o}, {Farr{\` a}s
  Casas}, {Federici}, {Fedorets}, {Fernique}, { Figueras}, {Filippi},
  {Findeisen}, { Fonti}, {Fraile}, {Fraser}, {Fr{\'e} zouls}, {Gai}, {Galleti},
  {Garabato}, {Garc{\'\i} a-Sedano}, {Garofalo}, { Garralda}, {Gavel},
  {Gavras}, {Gerssen}, {Geyer}, {Giacobbe}, {Gilmore}, { Girona}, {Giuffrida},
  {Glass}, {Gomes}, {Granvik}, {Gueguen}, {Guerrier}, {Guiraud}, {Guti{\'e}},
  {Haigron}, { Hatzidimitriou}, {Hauser}, {Haywood}, { Heiter}, {Helmi}, {Heu},
  {Hilger}, {Hobbs}, {Hofmann}, {Holland}, { Huckle}, {Hypki}, {Icardi},
  {Jan{\ss} en}, {Jevardat de Fombelle}, {Jonker}, {Juh{\'a}sz}, {Julbe}, {
  Karampelas}, {Kewley}, {Klar}, {Kochoska}, { Kohley}, {Kolenberg},
  {Kontizas}, { Kontizas }, {Koposov}, {Kordopatis}, { Kostrzewa-Rutkowska},
  {Koubsky}, {Lambert}, { Lanza}, {Lasne}, {Lavigne}, {Le Fustec}, {Le
  Poncin-Lafitte}, {Lebreton }, {Leccia}, {Leclerc}, {Lecoeur-Taibi},
  {Lenhardt}, { Leroux}, {Liao}, { Licata}, {Lindstr{\o}m}, {Lister},
  {Livanou}, {Lobel}, {L{\'o} pez}, { Managau}, {Mann}, {Mantelet}, { Marchal},
  {Marchant}, {Marconi}, { Marinoni}, {Marschalk{\'o}}, {Marshall}, {Martino},
  {Marton}, {Mary}, { Massari}, {Matijevi{\v{c} }}, {Mazeh}, { McMillan},
  {Messina}, { Michalik}, { Millar}, {Molina}, {Molinaro}, {Moln{ \'a}r},
  {Montegriffo}, {Mor}, { Morbidelli}, {Morel}, {Morris}, {Mulone}, {Muraveva},
  {Musella}, {Nelemans}, {Nicastro }, {Noval}, {O'Mullane}, {Ord{\'e}novic},
  {Ord{\'o}{\~n}ez-Blanco}, {Osborne}, {Pagani}, { Pagano}, {Pailler },
  {Palacin}, {Palaversa}, {Panahi}, {Pawlak}, {Piersimoni}, {Pineau },
  {Plachy}, {Plum}, {Poggio}, { Poujoulet}, {Pr{\v{s}}a}, {Pulone}, { Racero},
  {Ragaini}, {Rambaux}, { Ramos-Lerate}, { Regibo}, {Reyl{\'e}}, { Riclet},
  {Ripepi}, {Riva}, {Rivard}, {Rixon}, {Roegiers}, { Roelens}, {
  Romero-G{\'o}mez}, {Rowell}, {Royer}, { Ruiz-Dern}, {Sadowski}, {Sagrist{\`a}
  Sell{\'e} s}, {Sahlmann}, {Salgado}, {Salguero}, { Sanna}, {Santana-Ros},
  {Sarasso}, {Savietto}, {Schultheis}, {Sciacca}, {Segol}, {Segovia},
  {S{\'e}gransan}, {Shih}, {Siltala}, {Silva}, { Smart}, {Smith}, {Solano}, {
  Solitro}, {Sordo}, {Soria Nieto}, { Souchay}, {Spagna}, {Spoto}, {Stampa}, {
  Steele}, {Steidelm{\"u}ller}, { Stephenson}, {Stoev}, {Suess}, { Surdej},
  {Szabados}, {Szegedi-Elek}, { Tapiador}, {Taris}, { Tauran}, {Taylor},
  {Teixeira}, {Terrett}, {Teyssandier}, {Thuillot}, {Titarenko}, { Torra
  Clotet}, {Turon}, {Ulla}, {Utrilla}, { Uzzi}, {Vaillant}, {Valentini},
  {Valette }, {van Elteren}, {Van Hemelryck}, { Vaschetto}, {Vecchiato},
  {Veljanoski}, { Viala}, {Vicente }, {Vogt}, {von Essen}, {Voss}, {Votruba},
  {Voutsinas}, {Walmsley}, {Weiler}, { Wertz}, {Wevers}, {Wyrzykowski},
  {Yoldas}, {{\v{Z}} erjal}, {Ziaeepour}, {Zorec}, { Zschocke }, {Zucker},
  {Zurbach}, \& {Zwitter}}]{GaiaCollaboration2018b}
{Gaia Collaboration}, {Babusiaux}, C., {van Leeuwen}, F., {et~al.} 2018, \aap,
  616, A10

\bibitem[{{Gaia Collaboration} {et~al.}(2021{\natexlab{b}}){Gaia
  Collaboration}, {Brown}, {Vallenari}, { Prusti}, {de Bruijne}, { Babusiaux},
  { Biermann}, {Creevey}, { Evans}, {Eyer}, {Hutton}, {Jansen}, {Jordi},
  {Klioner}, {Lammers}, {Lindegren}, {Luri}, {Mignard}, { Panem}, {Pourbaix},
  {Randich}, { Sartoretti}, {Soubiran}, {Walton}, { Arenou}, {Bailer-Jones},
  {Bastian}, {Cropper}, {Drimmel}, {Katz}, { Lattanzi}, {van Leeuwen},
  {Bakker}, { Cacciari}, {Casta{\~n}eda}, {De Angeli}, {Ducourant},
  {Fabricius}, { Fouesneau}, {Fr{\'e}mat}, {Guerra}, {Guerrier}, { Guiraud},
  {Jean-Antoine Piccolo}, {Masana}, {Messineo}, {Mowlavi}, {Nicolas}, {
  Nienartowicz}, {Pailler}, {Panuzzo}, { Riclet}, {Roux}, {Seabroke}, {Sordo},
  {Tanga}, {Th{\'e}venin}, {Gracia-Abril}, {Portell}, {Teyssier}, {Altmann},
  {Andrae}, { Bellas-Velidis}, {Benson}, { Berthier}, {Blomme}, {Brugaletta}, {
  Burgess}, {Busso}, { Carry}, {Cellino }, {Cheek}, {Clementini}, {Damerdji},
  {Davidson}, {Delchambre}, { Dell'Oro}, {Fern{\'a}ndez-Hern{\'a}ndez},
  {Galluccio}, { Garc{\'\i}a-Lario}, {Garcia-Reinaldos}, {Gonz {\'a}lez-N
  {\'u}{\~n}ez}, {Gosset}, {Haigron}, { Halbwachs}, {Hambly}, {Harrison }, {
  Hatzidimitriou}, {Heiter}, { Hern{ \'a}ndez}, { Hestroffer}, {Hodgkin}, {
  Holl}, {Jan{\ss}en}, { Jevardat de Fombelle}, {Jordan}, { Krone-Martins}, {
  Lanzafame}, {L{\"o}ffler}, { Lorca}, { Manteiga}, {Marchal}, {Marrese}, {
  Moitinho}, {Mora}, {Muinonen}, {Osborne}, { Pancino}, {Pauwels}, {Petit}, {
  Recio-Blanco}, { Richards}, {Riello}, { Rimoldini}, {Robin}, {Roegiers}, {
  Rybizki}, {Sarro}, { Siopis}, {Smith}, { Sozzetti}, {Ulla}, { Utrilla}, {van
  Leeuwen}, {van Reeven}, { Abbas}, { Abreu Aramburu}, { Accart}, { Aerts}, {
  Aguado}, {Ajaj}, { Altavilla}, {{\'A} lvarez}, {{\'A}lvarez Cid-Fuentes },
  {Alves}, {Anderson }, {Anglada Varela}, { Antoja}, {Audard}, { Baines},
  {Baker}, {Balaguer-N{\'u}{\~n}ez}, {Balbinot}, { Balog}, {Barache},
  {Barbato}, {Barros}, {Barstow}, {Bartolom{\'e}}, { Bassilana}, {Bauchet },
  {Baudesson-Stella}, {Becciani}, { Bellazzini}, {Bernet}, {Bertone}, {
  Bianchi}, { Blanco-Cuaresma}, {Boch}, { Bombrun}, {Bossini}, { Bouquillon}, {
  Bragaglia}, {Bramante}, { Breedt}, { Bressan}, { Brouillet}, { Bucciarelli},
  {Burlacu}, {Busonero}, { Butkevich}, {Buzzi}, {Caffau}, { Cancelliere}, {C{\'
  a} novas}, {Cantat-Gaudin}, { Carballo}, {Carlucci}, {Carnerero}, {Carrasco
  }, {Casamiquela}, { Castellani}, { Castro-Ginard}, {Castro Sampol}, {Chaoul},
  { Charlot}, {Chemin}, { Chiavassa}, {Cioni}, {Comoretto}, { Cooper},
  {Cornez}, { Cowell}, { Crifo}, { Crosta}, {Crowley}, { Dafonte}, { Dapergolas
  }, {David}, {David}, { de Laverny}, {De Luise}, {De March}, {De Ridder }, {de
  Souza}, {de Teodoro}, {de Torres}, {del Peloso}, {del Pozo}, { Delbo},
  {Delgado}, {Delgado}, { Delisle}, {Di Matteo}, {Diakite}, { Diener},
  {Distefano}, {Dolding}, { Eappachen}, {Edvardsson}, {Enke}, { Esquej},
  {Fabre}, {Fabrizio}, {Faigler}, {Fedorets }, { Fernique}, {Fienga},
  {Figueras}, { Fouron}, {Fragkoudi}, {Fraile}, {Franke}, {Gai}, {Garabato},
  {Garcia-Gutierrez}, {Garc{\' \i} a-Torres}, {Garofalo}, {Gavras}, {Gerlach},
  { Geyer}, {Giacobbe}, {Gilmore}, {Girona}, {Giuffrida}, {Gomel}, {Gomez}, {
  Gonzalez-Santamaria}, {Gonz{\'a}lez-Vidal}, {Granvik },
  {Guti{\'e}rrez-S{\'a}nchez}, {Guy}, { Hauser}, {Haywood}, {Helmi}, {Hidalgo},
  {Hilger}, {H{\l}adczuk}, { Hobbs}, {Holland}, {Huckle}, { Jasniewicz},
  {Jonker}, { Juaristi Campillo}, {Julbe}, {Karbevska}, { Kervella}, {Khanna},
  {Kochoska}, {Kontizas}, { Kordopatis}, {Korn}, {Kostrzewa-Rutkowska},
  {Kruszy{\'n}ska}, {Lambert}, {Lanza}, { Lasne}, {Le Campion}, {Le Fustec}, {
  Lebreton }, {Lebzelter}, {Leccia}, {Leclerc}, { Lecoeur-Taibi}, {Liao},
  {Licata}, {Lindstr {\o}m}, {Lister}, {Livanou}, {Lobel}, {Madrero Pardo},
  {Managau}, {Mann}, { Marchant }, {Marconi}, {Marcos Santos}, { Marinoni},
  {Marocco}, {Marshall}, { Martin Polo}, {Mart{\'\i}n-Fleitas}, {Masip}, {
  Massari}, {Mastrobuono-Battisti}, { Mazeh}, { McMillan}, {Messina}, {
  Michalik}, {Millar}, {Mints}, {Molina }, {Molinaro}, {Moln{ \'a}r}, {
  Montegriffo}, {Mor}, {Morbidelli}, { Morel}, {Morris}, {Mulone}, {Munoz},
  {Muraveva}, {Murphy}, {Musella}, {Noval}, {Ord{\'e}novic}, {Orr{\`u}},
  {Osinde}, { Pagani}, {Pagano}, { Palaversa}, {Palicio}, {Panahi}, { Pawlak},
  {Pe{\~n}alosa Esteller}, {Penttil{\"a} }, {Piersimoni}, {Pineau}, { Plachy},
  {Plum}, {Poggio}, {Poretti}, { Poujoulet}, {Pr{\v{s}}a}, {Pulone}, { Racero},
  {Ragaini}, {Rainer}, { Raiteri}, {Rambaux}, {Ramos}, { Ramos-Lerate}, {Re
  Fiorentin}, {Regibo}, {Reyl{\'e}}, {Ripepi}, {Riva}, {Rixon}, {Robichon},
  {Robin}, {Roelens}, { Rohrbasser}, {Romero-G{\'o}mez}, { Rowell}, {Royer},
  {Rybicki}, {Sadowski}, { Sagrist{\`a} Sell{\'e}s}, {Sahlmann}, {Salgado},
  {Salguero}, {Samaras}, {Sanchez Gimenez}, {Sanna}, { Santove{\~n}a}, {
  Sarasso}, { Schultheis}, { Sciacca}, { Segol}, {Segovia}, {S{\'e}gransan},
  {Semeux}, {Shahaf}, { Siddiqui}, { Siebert}, {Siltala}, {Slezak}, {Smart},
  {Solano}, {Solitro}, {Souami}, {Souchay}, {Spagna}, {Spoto}, { Steele},
  {Steidelm{\" u}ller}, {Stephenson}, {S{\"u }veges}, { Szabados}, {
  Szegedi-Elek}, {Taris}, {Tauran}, { Taylor}, { Teixeira}, { Thuillot}, {
  Tonello}, { Torra}, {Torra}, {Turon}, { Unger}, {Vaillant}, {van Dillen},
  {Vanel}, { Vecchiato}, {Viala}, {Vicente }, { Voutsinas}, {Weiler}, {Wevers},
  { Wyrzykowski}, { Yoldas}, {Yvard}, { Zhao}, {Zorec}, { Zucker}, { Zurbach},
  \& {Zwitter}}]{GaiaCollaboration:2021d}
{Gaia Collaboration}, {Brown}, A.~G.~A., {Vallenari}, A., {et~al.}
  2021{\natexlab{b}}, \aap, 650, C3

\bibitem[{{Geisler} {et~al.}(1995){Geisler}, {Piatti}, {Claria}, \& {
  Minniti}}]{Geisler:1995}
{Geisler}, D., {Piatti}, A.~E., {Claria}, J.~J., \& { Minniti}, D. 1995, \aj,
  109, 605

\bibitem[{{Green}(2018)}]{Green:2018}
{Green}, G. 2018, The Journal of Open Source Software, 3, 695

\bibitem[{{Grillmair} \& {Dionatos}(2006)}]{Grillmair:2006}
{Grillmair}, C.~J. \& {Dionatos}, O. 2006, \apj, 643, L17

\bibitem[{{Hansen} {et~al.}(2020){Hansen}, {Riley}, {Strigari}, { Marshall},
  {Ferguson}, {Zepeda}, \& {Sneden}}]{Hansen:2020}
{Hansen}, T.~T., {Riley}, A.~H., {Strigari}, L.~E., {et~al.} 2020, \apj, 901,
  23

\bibitem[{{Harris}(1996)}]{Harris:1996}
{Harris}, W.~E. 1996, \aj, 112, 1487

\bibitem[{{Helmi}(2020)}]{Helmi:2020}
{Helmi}, A. 2020, \araa, 58, 205

\bibitem[{{Helmi} \& {White}(1999)}]{Helmi:1999a}
{Helmi}, A. \& {White}, S. D.~M. 1999, \mnras, 307, 495

\bibitem[{{Hunter}(2007)}]{mpl}
{Hunter}, J.~D. 2007, Computing in Science and Engineering, 9, 90

\bibitem[{{Ibata} {et~al.}(2021){Ibata}, {Malhan}, {Martin}, {Aubert},
  {Famaey}, {Bianchini}, {Monari}, {Siebert}, {Thomas }, {Bellazzini},
  {Bonifacio}, {Caffau}, \& {Renaud}}]{Ibata:2021}
{Ibata}, R., {Malhan}, K., {Martin}, N., {et~al.} 2021, \apj, 914, 123

\bibitem[{{Ibata} {et~al.}(1994){Ibata}, {Gilmore}, \& {Irwin}}]{Ibata:1994}
{Ibata}, R.~A., {Gilmore}, G., \& {Irwin}, M.~J. 1994, \nat, 370, 194

\bibitem[{{Ibata} {et~al.}(2019){Ibata}, {Malhan}, \& {Martin}}]{Ibata:2019}
{Ibata}, R.~A., {Malhan}, K., \& {Martin}, N.~F. 2019, \apj, 872, 152

\bibitem[{{Ibata} {et~al.}(2018){Ibata}, {Malhan}, {Martin}, \&
  {Starkenburg}}]{Ibata:2018}
{Ibata}, R.~A., {Malhan}, K., {Martin}, N.~F., \& {Starkenburg}, E. 2018, \apj,
  865, 85

\bibitem[{Jones {et~al.}(2001--)Jones, Oliphant, Peterson, {et~al.}}]{scipy}
Jones, E., Oliphant, T., Peterson, P., {et~al.} 2001--, {SciPy}: Open source
  scientific tools for {Python}

\bibitem[{{Koposov} {et~al.}(2010){Koposov}, {Rix}, \& {Hogg}}]{Koposov:2010}
{Koposov}, S.~E., {Rix}, H.-W., \& {Hogg}, D.~W. 2010, \apj, 712, 260

\bibitem[{{Koppelman} {et~al.}(2019){Koppelman}, {Helmi}, {Massari},
  {Price-Whelan}, \& {Starkenburg}}]{Koppelman:2019a}
{Koppelman}, H.~H., {Helmi}, A., {Massari}, D., {Price-Whelan}, A.~M., \&
  {Starkenburg}, T.~K. 2019, \aap, 631, L9

\bibitem[{{K{\"u}pper} {et~al.}(2015){K{\"u}pper}, {Balbinot}, { Bonaca},
  {Johnston}, {Hogg}, {Kroupa}, \& {Santiago}}]{Kupper:2015}
{K{\"u}pper}, A. H.~W., {Balbinot}, E., { Bonaca}, A., {et~al.} 2015, \apj,
  803, 80

\bibitem[{{Li} {et~al.}(2018{\natexlab{a}}){Li}, {Yanny}, \& {Wu}}]{Li:2018}
{Li}, G.-W., {Yanny}, B., \& {Wu}, Y. 2018{\natexlab{a}}, \apj, 869, 122

\bibitem[{{Li} {et~al.}(2022){Li}, {Aoki}, {Matsuno}, { Xing}, {Suda},
  {Tominaga}, {Chen}, {Honda}, {Ishigaki}, {Shi}, {Zhao}, \& {Zhao}}]{Li:2022}
{Li}, H., {Aoki}, W., {Matsuno}, T., {et~al.} 2022, \apj, 931, 147

\bibitem[{{Li} {et~al.}(2018{\natexlab{b}}){Li}, {Tan}, \& {Zhao}}]{Li:2018VMP}
{Li}, H., {Tan}, K., \& {Zhao}, G. 2018{\natexlab{b}}, \apjs, 238, 16

\bibitem[{{Lindegren} {et~al.}(2021){Lindegren}, {Klioner}, {Hern{\'a}ndez}, {
  Bombrun}, {Ramos-Lerate}, {Steidelm{\"u} ller}, { Bastian}, {Biermann}, {de
  Torres}, {Gerlach}, {Geyer}, {Hilger}, {Hobbs}, {Lammers}, {McMillan},
  {Stephenson}, {Casta{\~n}eda}, {Davidson }, {Fabricius}, {Gracia-Abril}, {
  Portell}, {Rowell}, {Teyssier}, {Torra}, {Bartolom{\'e}}, {Clotet},
  {Garralda}, {Gonz{\'a }lez-Vidal}, {Torra}, { Abbas}, {Altmann}, {Anglada
  Varela}, { Balaguer-N{\'u}{\~n}ez}, {Balog }, {Barache}, {Becciani},
  {Bernet}, { Bertone}, {Bianchi}, {Bouquillon}, {Brown}, {Bucciarelli},
  {Busonero}, {Butkevich}, {Buzzi}, {Cancelliere}, {Carlucci}, {Charlot },
  {Cioni}, {Crosta}, { Crowley}, { del Peloso}, {del Pozo}, {Drimmel}, {Esquej
  }, {Fienga}, {Fraile}, {Gai}, { Garcia-Reinaldos}, {Guerra}, {Hambly}, {
  Hauser}, {Jan{\ss}en}, {Jordan}, { Kostrzewa-Rutkowska}, {Lattanzi }, {Liao},
  { Licata}, {Lister}, {L{\"o}ffler}, {Marchant}, {Masip}, {Mignard}, {Mints},
  {Molina}, {Mora}, {Morbidelli}, {Murphy}, { Pagani}, {Panuzzo}, {Pe{\~n}alosa
  Esteller}, { Poggio}, {Re Fiorentin}, {Riva}, { Sagrist{\`a} Sell{\'e}s},
  {Sanchez Gimenez}, { Sarasso}, { Sciacca}, {Siddiqui}, { Smart}, {Souami},
  {Spagna}, {Steele}, {Taris}, {Utrilla}, {van Reeven}, \&
  {Vecchiato}}]{Lindegren:2021}
{Lindegren}, L., {Klioner}, S.~A., {Hern{\'a}ndez}, J., {et~al.} 2021, \aap,
  649, A2

\bibitem[{{Malhan} \& {Ibata}(2018)}]{Malhan:2018}
{Malhan}, K. \& {Ibata}, R.~A. 2018, \mnras, 477, 4063

\bibitem[{{Mar{\'\i}n-Franch} {et~al.}(2009){Mar{\'\i}n-Franch}, {Aparicio}, {
  Piotto}, {Rosenberg}, {Chaboyer}, { Sarajedini}, {Siegel}, { Anderson}, {
  Bedin }, {Dotter}, {Hempel}, {King}, {Majewski}, { Milone}, {Paust}, \&
  {Reid}}]{Marin-Franch:2009}
{Mar{\'\i}n-Franch}, A., {Aparicio}, A., { Piotto}, G., {et~al.} 2009, \apj,
  694, 1498

\bibitem[{{Martin} {et~al.}(2022){Martin}, {Ibata}, { Starkenburg}, {Yuan},
  {Malhan}, {Bellazzini}, { Viswanathan}, {Aguado }, {Arentsen}, { Bonifacio},
  {Carlberg}, {Gonz{\'a}lez Hern{\'a} ndez}, {Hill}, {Jablonka}, { Kordopatis},
  {Lardo}, {McConnachie}, {Navarro}, {S{\'a}nchez-Janssen}, { Sestito},
  {Thomas}, {Venn}, {Vitali}, \& {Voggel}}]{Martin:2022}
{Martin}, N.~F., {Ibata}, R.~A., { Starkenburg}, E., {et~al.} 2022, \mnras,
  516, 5331

\bibitem[{{Massari} {et~al.}(2019){Massari}, {Koppelman}, \&
  {Helmi}}]{Massari:2019}
{Massari}, D., {Koppelman}, H.~H., \& {Helmi}, A. 2019, \aap, 630, L4

\bibitem[{{Mateu} {et~al.}(2018){Mateu}, {Read}, \& {Kawata}}]{galstreams}
{Mateu}, C., {Read}, J.~I., \& {Kawata}, D. 2018, \mnras, 474, 4112

\bibitem[{{Mikkola} {et~al.}(2023){Mikkola}, {McMillan}, \&
  {Hobbs}}]{Mikkola:2023}
{Mikkola}, D., {McMillan}, P.~J., \& {Hobbs}, D. 2023, \mnras, 519, 1989

\bibitem[{{Myeong} {et~al.}(2019){Myeong}, {Vasiliev}, {Iorio}, { Evans}, \&
  {Belokurov}}]{Myeong:2019}
{Myeong}, G.~C., {Vasiliev}, E., {Iorio}, G., { Evans}, N.~W., \& {Belokurov},
  V. 2019, \mnras, 488, 1235

\bibitem[{{Odenkirchen} {et~al.}(2001){Odenkirchen}, {Grebel}, {Rockosi},
  {Dehnen}, {Ibata}, { Rix}, {Stolte}, {Wolf}, {Anderson}, {Bahcall}, {
  Brinkmann}, {Csabai}, {Hennessy}, {Hindsley}, {Ivezi{\'c}}, {Lupton}, {Munn},
  { Pier}, {Stoughton}, \& {York}}]{Odenkirchen:2001}
{Odenkirchen}, M., {Grebel}, E.~K., {Rockosi}, C.~M., {et~al.} 2001, \apj, 548,
  L165

\bibitem[{P\'erez \& Granger(2007)}]{ipython}
P\'erez, F. \& Granger, B.~E. 2007, Computing in Science and Engineering, 9, 21

\bibitem[{{Price-Whelan}(2017)}]{gala}
{Price-Whelan}, A.~M. 2017, The Journal of Open Source Software, 2, 388

\bibitem[{{Riley} \& {Strigari}(2020)}]{Riley:2020}
{Riley}, A.~H. \& {Strigari}, L.~E. 2020, \mnras, 494, 983

\bibitem[{{Schlegel} {et~al.}(1998){Schlegel}, {Finkbeiner}, \&
  {Davis}}]{Schlegel:1998}
{Schlegel}, D.~J., {Finkbeiner}, D.~P., \& {Davis}, M. 1998, \apj, 500, 525

\bibitem[{{Shipp} {et~al.}(2018){Shipp}, {Drlica-Wagner}, {Balbinot}, {
  Ferguson }, {Erkal}, {Li}, {Bechtol}, { Belokurov}, {Buncher}, {Carollo},
  {Carrasco Kind }, {Kuehn}, {Marshall}, {Pace}, { Rykoff}, {Sevilla-Noarbe },
  {Sheldon}, { Strigari}, {Vivas}, {Yanny}, {Zenteno}, {Abbott}, {Abdalla},
  {Allam}, { Avila}, { Bertin}, {Brooks}, {Burke}, { Carretero}, {Castander},
  {Cawthon}, { Crocce}, {Cunha}, {D'Andrea}, {da Costa}, {Davis}, {De Vicente},
  {Desai }, { Diehl}, {Doel}, {Evrard}, {Flaugher}, {Fosalba}, {Frieman}, {
  Garc{\'\i}a-Bellido}, {Gaztanaga}, {Gerdes}, {Gruen}, {Gruendl }, {
  Gschwend}, {Gutierrez}, {Hoyle}, {James}, { Johnson}, {Krause}, {Kuropatkin
  }, { Lahav }, {Lin}, {Maia}, {March }, {Martini }, {Menanteau}, {Miller},
  {Miquel}, { Nichol}, {Plazas}, {Romer}, {Sako}, {Sanchez}, { Scarpine }, {
  Schindler}, { Schubnell}, {Smith}, {Smith}, {Sobreira}, { Suchyta},
  {Swanson}, { Tarle}, { Thomas }, {Tucker}, {Walker}, { Wechsler}, \& {the DES
  Collaboration}}]{Shipp:2018}
{Shipp}, N., {Drlica-Wagner}, A., {Balbinot}, E., {et~al.} 2018, ArXiv
  e-prints, 862, 114

\bibitem[{{Shipp} {et~al.}(2021){Shipp}, {Erkal}, {Drlica-Wagner}, {Li},
  {Pace}, {Koposov}, {Cullinane }, {Da Costa}, {Ji}, {Kuehn }, {Lewis},
  {Mackey}, {Simpson}, {Wan}, {Zucker}, { Bland-Hawthorn}, {Ferguson},
  {Lilleengen}, \& {S5 Collaboration}}]{Shipp:2021}
{Shipp}, N., {Erkal}, D., {Drlica-Wagner}, A., {et~al.} 2021, arXiv e-prints,
  arXiv:2107.13004

\bibitem[{{Tenachi} {et~al.}(2022){Tenachi}, {Oria}, {Ibata}, {Famaey}, {Yuan},
  {Arentsen}, { Martin}, \& {Viswanathan}}]{Tenachi:2022}
{Tenachi}, W., {Oria}, P.-A., {Ibata}, R., {et~al.} 2022, \apjl, 935, L22

\bibitem[{{Vasiliev} \& {Baumgardt}(2021)}]{Vasiliev21}
{Vasiliev}, E. \& {Baumgardt}, H. 2021, \mnras, 505, 5978

\bibitem[{Walt {et~al.}(2011)Walt, Colbert, \& Varoquaux}]{numpy}
Walt, S. v.~d., Colbert, S.~C., \& Varoquaux, G. 2011, Computing in Science and
  Engg., 13, 22

\bibitem[{{Wan} {et~al.}(2020){Wan}, {Lewis}, {Li}, { Simpson}, {Martell},
  {Zucker}, { Mould }, {Erkal}, { Pace}, { Mackey}, {Ji}, {Koposov}, { Kuehn},
  {Shipp}, {Balbinot}, { Bland-Hawthorn}, {Casey }, {Da Costa}, {Kafle},
  {Sharma}, \& {De Silva}}]{Wan:2020}
{Wan}, Z., {Lewis}, G.~F., {Li}, T.~S., {et~al.} 2020, \nat, 583, 768

\bibitem[{{Willman} \& {Strader}(2012)}]{Willman:2012}
{Willman}, B. \& {Strader}, J. 2012, \aj, 144, 76

\bibitem[{{Yanny} {et~al.}(2009){Yanny}, {Rockosi}, {Newberg}, {Knapp},
  {Adelman-McCarthy}, {Alcorn}, {Allam}, {Allende Prieto}, {An}, {Anderson},
  {Anderson}, { Bailer-Jones}, { Bastian}, {Beers}, {Bell}, {Belokurov},
  {Bizyaev}, {Blythe}, {Bochanski}, {Boroski}, { Brinchmann}, {Brinkmann},
  {Brewington}, {Carey}, {Cudworth}, {Evans}, { Evans}, {Gates}, {G{\"a}
  nsicke}, { Gillespie}, {Gilmore}, {Nebot Gomez-Moran}, {Grebel}, {
  Greenwell}, {Gunn}, { Jordan}, {Jordan}, {Harding}, {Harris}, {Hendry},
  {Holder}, {Ivans}, {Ivezi{\v{c}}}, {Jester}, { Johnson}, {Kent}, {Kleinman},
  {Kniazev}, {Krzesinski}, { Kron}, { Kuropatkin}, {Lebedeva}, {Lee}, {French
  Leger}, {L{ \'e}pine}, { Levine}, {Lin}, {Long}, {Loomis}, { Lupton},
  {Malanushenko}, { Malanushenko}, {Margon}, {Martinez-Delgado}, { McGehee},
  {Monet}, {Morrison}, {Munn}, {Neilsen}, { Nitta}, { Norris}, {Oravetz},
  {Owen}, { Padmanabhan}, {Pan}, {Peterson}, {Pier }, {Platson}, {Re
  Fiorentin}, { Richards}, {Rix}, { Schlegel}, { Schneider}, {Schreiber}, {
  Schwope}, {Sibley}, {Simmons}, {Snedden}, {Allyn Smith}, {Stark}, {Stauffer},
  {Steinmetz}, { Stoughton}, {SubbaRao}, {Szalay}, { Szkody}, { Thakar},
  {Sivarani}, {Tucker}, {Uomoto}, { Vanden Berk}, {Vidrih}, {Wadadekar}, {
  Watters}, {Wilhelm}, {Wyse}, { Yarger}, \& {Zucker}}]{Yanny:2009}
{Yanny}, B., {Rockosi}, C., {Newberg}, H.~J., {et~al.} 2009, \aj, 137, 4377

\bibitem[{{Yuan} {et~al.}(2020){Yuan}, {Myeong}, {Beers}, {Evans}, {Lee},
  {Banerjee}, {Gudin}, {Hattori}, {Li}, {Matsuno}, {Placco}, {Smith},
  {Whitten}, \& {Zhao}}]{Yuan:2020}
{Yuan}, Z., {Myeong}, G.~C., {Beers}, T.~C., {et~al.} 2020, \apj, 891, 39

\bibitem[{{Zinn} \& {West}(1984)}]{Zinn:1984}
{Zinn}, R. \& {West}, M.~J. 1984, \apjs, 55, 45

\end{thebibliography}

\begin{appendix}
	\section{List of members}

	Here we list all the members of ED-2 used in this work. We give only their
	Gaia DR3 \texttt{source\_id} values, and we note that we also include the
	velocity outliers that were removed in Sect \ref{sec:6d}.

	\begin{table}
		\footnotesize
		\centering
		\caption{Gaia DR3 \texttt{source\_id} for ED-2 members used in this work}
		\label{tab:members}
		\begin{tabular}{c}
			\hline
			\texttt{source\_id} \\
			\hline
			\hline
			759407156314566784  \\
			779616592350301952  \\
			1298276602498467072 \\
			1355410284693698304 \\
			1473531196822925696 \\
			1483840527082796160 \\
			1504043950164231040 \\
			1505084294322771200 \\
			1571514072452538752 \\
			1574800645851240192 \\
			1577319631286280960 \\
			1578782394069199488 \\
			1607956973038473600 \\
			1613368013356040192 \\
			1665210536361012480 \\
			1693676308288915712 \\
			1698151522476505344 \\
			1828751586656157952 \\
			1835597214774359424 \\
			2110672655835482496 \\
			3473979147705211776 \\
			3549718318990080896 \\
			3592103255289580800 \\
			3624621311681013504 \\
			3723617593434390656 \\
			3757312745743087232 \\
			3786501309126931328 \\
			3869876996687740032 \\
			3915925991063287808 \\
			3941060277118782208 \\
			3951060094855048960 \\
			3963873700285600128 \\
			3973419660238121728 \\
			3992080193627189632 \\
			4008274384997266560 \\
			4014326268795251712 \\
			4025946010756238464 \\
			4154818772254357376 \\
			4172609385769404032 \\
			4245522468554091904 \\
			4479226310758314496 \\
			4532592619428218624 \\
			5444820480268295424 \\
			5991844282681283712 \\
			6632335060231088896 \\
			6646097819069706624 \\
			6746114585056265600 \\
			6747065215934660608 \\
			\hline
		\end{tabular}
	\end{table}

\end{appendix}

%
%
%
\end{document}